\documentclass[11pt]{article}

\PassOptionsToPackage{hyperfootnotes=false}{hyperref}
\usepackage[preprint]{acl}

\usepackage{times}
\usepackage{latexsym}

\usepackage[T1]{fontenc}
\usepackage[utf8]{inputenc}

\usepackage{microtype}

\usepackage{inconsolata}

\usepackage{graphicx}

\usepackage{twemojis}
\usepackage{amsmath}
\usepackage{amsfonts, amssymb}
\usepackage{array}
\usepackage{booktabs}
\usepackage{needspace}
\title{MoE-CORE: Coordinated Expert Offloading and Residency for Memory-Constrained MoE Inference}

\author{
 \textbf{Ke Yang\textsuperscript{3}}\thanks{The two authors contributed equally to this work.},
 \textbf{Yongji Gao\textsuperscript{1,2}}\footnotemark[1],
 \textbf{Xushi Li\textsuperscript{1,2}}\thanks{Corresponding author},
 \textbf{Kui Luo\textsuperscript{1}},
 \textbf{Sicheng Zhang\textsuperscript{1,2}},
 \textbf{Tianming Zhou\textsuperscript{1,2}},\\
 \textbf{Keyi Liu\textsuperscript{1}},
 \textbf{Shufang Lu\textsuperscript{1}},
 \textbf{Aoxuan Chen\textsuperscript{1}},
 \textbf{Jie Meng\textsuperscript{1}},
 \textbf{Jingchun Gao\textsuperscript{1,2}},
 \textbf{Dan Li\textsuperscript{1,2}},
 \textbf{Xinkai You\textsuperscript{1}},\\
 \textbf{Dan Li\textsuperscript{1}},
 \textbf{Zhixiang Xia\textsuperscript{1,2}},
 \textbf{Yan Shi\textsuperscript{4}},
 \textbf{Yang Liu\textsuperscript{1}},
 \textbf{Yanjia Zeng\textsuperscript{1}},
 \textbf{Liangjun Feng\textsuperscript{1,2}}
\\
 \textsuperscript{1}Huawei Technologies Co. Ltd\\
 \textsuperscript{2}Huawei Ascend Computing Technology Development\\
 \textsuperscript{3}University of Chinese Academy of Sciences\\
 \textsuperscript{4}Shanghai Jiao Tong University
 \\
 {\texttt{yangke@iie.ac.cn \quad keyi.liu@h-partners.com \quad amy\_sy@sjtu.edu.cn}}\\
 {\texttt{\{gaoyongji2, lixushi, luokui, zhangsicheng5, zhoutianming2,}}\\
 {\texttt{shufang.lu, chenaoxuan1, mengjie48, gaojingchun1, lidan276, youxinkai,}}\\
 {\texttt{xiazhixiang1, yang.liu8, zengyanjia, fengliangjun\}@huawei.com}}
}

\begin{document}
\maketitle
\begin{abstract}
Sparse expert activation reduces MoE models' computation, yet expert weights can exceed limited device memory. Offloading makes inference feasible on a compact AI appliance but exposes host-to-device transfers to the inference path. We present MoE-CORE, a system that coordinates expert offloading and residency for memory-constrained MoE inference. It stages complete expert layers in alternating buffers during prefill. During decode, it combines nonuniform layer-wise cache capacity, domain-informed initialization, routing-history-aware replacement, and cross-layer prefetching. The main configuration executes router-selected experts exactly; an optional score-based substitution path handles eligible low-score misses. The main comparison uses 1K- and 128-token output caps for MoE-CORE and vLLM Prefetch, respectively. Across five workloads per model, MoE-CORE records a mean time per output token (TPOT) of 38.0--44.8 ms versus 1268.9--1269.1 ms for the evaluated vLLM Prefetch configuration on DeepSeek-V4-Flash-W4A8; the corresponding values on GLM-5.2-W4A8C8 are 206.6--220.5 and 5941.5--5941.8 ms. Under an 84-GB NPU-memory cap, the best measured DeepSeek GSM8K configuration achieves a TPOT of 21.5 ms with approximate expert substitution and multi-token prediction (MTP) at depth 2. These results support coordinated expert residency and transfer scheduling under a device-memory constraint. The code is 
\href{https://github.com/LookAround0301/vllm-ascend/tree/moe_offload_v5.0/vllm_ascend/expert_offload}{here}.
\end{abstract}

\section{Introduction}

Mixture-of-Experts (MoE) architectures \cite{shazeer2017outrageouslylargeneuralnetworks} have become an important approach to scaling the capacity of large language models (LLMs) without proportionally increasing per-token computation. By routing each token to activate only a small subset of experts, MoE models offer a favorable computation-to-capacity ratio and are therefore attractive for resource-constrained local and edge inference \cite{yi2025edgemoeempoweringsparselarge}. Sparse activation, however, reduces computation rather than the amount of model weights that must remain accessible. Recent model families make this distinction increasingly consequential: DeepSeek-V4-Flash and DeepSeek-V4-Pro \cite{deepseekai2026deepseekv4} contain 284B and 1.6T total parameters while activating 13B and 49B parameters per token, respectively; GLM-5.2 \cite{glm5team2026glm5vibecodingagentic} contains 744B total parameters with 40B active; and Kimi K3 scales to 2.8T total parameters with 104B active \cite{kimik3_2026}. These models also increasingly target Agent workloads such as long-horizon coding, tool use, and knowledge work. Deploying them as local Agent services therefore requires attention not only to sparse computation, but also to the storage capacity and memory cost of the full expert set.

This memory burden is particularly acute on compact, single-node AI appliances. These appliances integrate a host processor and memory with one or a small number of GPUs or NPUs, providing enterprises with a practical on-premises platform for deploying AI services close to private data while retaining local control over the serving environment. However, the device-memory capacity of these systems is under increasing pressure from two directions: Growing MoE parameter counts increase the volume of weights, while long-context Agent tasks expand the KV cache with conversation histories, retrieved documents, tool outputs, and intermediate task state. Thus, these demands can prevent the complete set of model weights from residing in device memory, even though an AI appliance provides substantially more resources than a mobile or embedded edge device. Deploying LLMs on such systems therefore requires placing some of the model weights in host memory.

Expert offloading makes such deployments feasible by keeping some routed expert weights in host memory and transferring them to the device memory (GPUs or NPUs) when needed. 
Because expert weights account for a large fraction of an MoE model's parameters while each token activates only a small subset of experts, retaining inactive experts in host memory can substantially increase the model size supported by a fixed device-memory budget.
This technique reduces device-memory demand, but loading a selected expert after routing places host-to-device (H2D) communication on the inference critical path. Existing techniques for reducing this cost can be grouped into four categories. (i) \textbf{Activation-locality methods} use routing traces or semantic history to identify experts that are likely to be reused and retain them in a device cache, as in MoE-Infinity and FineMoE \cite{xue2025moeinfinityefficientmoeinference,yu2025taminglatencymemorytradeoffmoebased}. (ii) \textbf{Auxiliary-predictor methods} train a separate lightweight model to predict expert activations before the original router is reached. SiDA-MoE \cite{MLSYS2024_698cfaf7} uses an offline-trained hash predictor, while ProMoE trains predictors for individual layers \cite{song2025promoefastmoebasedllm}. (iii) \textbf{Cross-layer and phase-aware methods} obtain expert information from later gates or reorganize transfers according to layer and inference phase. Fate \cite{Fate2026Fang} and AdapMoE \cite{Zhong_2024} use cross-layer gating with layer-aware expert management, whereas DuoServe-MoE \cite{zhang2026duoservemoedualphaseexpertprefetch} separates prefill and decode transfer pipelines. (iv) \textbf{Expert-substitution methods}, including SMoE and BuddyMoE \cite{zhu2026smoealgorithmsystemcodesignpushing,wang2025buddymoeexploitingexpertredundancy}, avoid selected transfers in eligible cases by executing an already resident alternative expert. Together, these systems show that offloading latency can be addressed through caching, prediction, scheduling, and fallback mechanisms. However, these methods have been evaluated on short-context workloads, leaving the interaction between long contexts, expert residency, and offloading latency insufficiently characterized.

For memory-constrained MoE inference on a compact AI appliance, however, these mechanisms must operate as parts of a coordinated runtime rather than as isolated policies. This requirement raises three design questions:

\twemoji{gear} \textit{Does equal per-layer cache allocation yield comparable expert hit rates across MoE layers?}

\twemoji{bar_chart} \textit{Do expert popularity rankings differ across domain-specific workloads?}

\twemoji{arrows_counterclockwise} \textit{Are routing scores near the Top-\(k\) boundary close enough to motivate a resident alternative for an eligible low-score cache miss?}

To address these questions, we present MoE-CORE, a coordinated expert-offloading system for memory-constrained MoE inference on compact, single-node AI appliances.

MoE-CORE assigns prefill and decode to different expert-offloading paths and reserves a separate device-memory region for each phase. For prefills, it stages the complete routed-expert weights of two consecutive MoE layers in two alternating buffers, overlapping the transfer of the next layer with computation of the current layer. Decode is the primary optimization target because it activates a sparse expert set at each iteration, yet an uncovered expert transfer can directly delay token generation.

During decoding, MoE-CORE first distributes the expert cache budget across all MoE layers according to a dynamic layer-wise cache allocator. Given a workload-domain hint from the serving application, an offline profile initializes each layer's cache with the corresponding hot experts. A routing-history-aware online cache then adapts this initial placement to the routing behavior observed during the current session. Because expert residency alone cannot eliminate all misses, a cross-layer prefetcher evaluates the next layer's router early and overlaps predicted expert transfers with useful computation. For residual misses, high-score experts remain on the exact loading path, whereas only eligible low-score experts may use resident substitutes selected by routing-score proximity.

Relative to Load on Demand (LoD), which allocates static per-layer expert caches only for decode and synchronously loads non-resident selected experts, MoE-CORE achieves mean time per output token (TPOT) speedups of up to \(1.75\times\) on DeepSeek-V4-Flash-W4A8 and \(1.15\times\) on GLM-5.2-W4A8C8. MoE-CORE also records lower TPOT than the evaluated vLLM Prefetch configurations. Its best measured configuration achieves 21.5 ms TPOT with multi-token prediction (MTP) at depth 2.

Our contributions are as follows:
\begin{itemize}
    \item We characterize layer-wise differences in cache coverage and workload-dependent expert rankings under a fixed cache configuration, and examine routing scores near the Top-\(k\) activation boundary.
    \item We design phase-specific expert-offloading paths for prefill and decode within a shared device-memory budget.
    \item We coordinate layer-aware cache allocation, domain-informed initialization, online cache adaptation, and cross-layer prefetching; optional score-based substitution handles eligible low-score residual misses while high-score misses retain exact loading.
    \item We demonstrate MoE-CORE's effectiveness on five workloads with DeepSeek-V4-Flash-W4A8 and GLM-5.2-W4A8C8 on their respective AI-appliance platforms. It records lower TPOT than both LoD and the evaluated vLLM Prefetch configurations, while ablations isolate the effects of its cache-management and prefetching mechanisms.
\end{itemize}

\section{Background and Motivation}

\subsection{MoE Inference}\label{subsec::moe}
A Mixture-of-Experts (MoE) model replaces the dense feed-forward network in selected Transformer layers with multiple expert feed-forward networks (FFNs) and a router. For token \(t\) in MoE layer \(l\), the router scores all \(N_l\) routed experts from the hidden state \(h_{l,t}\) and selects the top-\(k_l\):
\begin{equation}
\mathbf{r}_{l,t}=R_l(h_{l,t}),\
\mathcal{S}_{l,t}=\operatorname{TopK}\!\left(\mathbf{r}_{l,t},k_l\right),
\end{equation}
where \(R_l\) is the router, \(\mathbf{r}_{l,t}\) contains the expert scores, and \(\mathcal{S}_{l,t}\) is the set of selected expert indices. The routed MoE output is
\begin{equation}
    m_{l,t}^{\mathrm{routed}}=\sum_{e\in\mathcal{S}_{l,t}}\alpha_{l,t,e}E_{l,e}(h_{l,t}),
\end{equation}
where \(E_{l,e}\) is the FFN of expert \(e\), and \(\alpha_{l,t,e}\) is its normalized routing weight. Because \(\mathcal{S}_{l,t}\) depends on the hidden state, expert selection can vary across tokens and layers. Each token evaluates only \(k_l\) experts, but the runtime must provide access to all \(N_l\) expert weights \cite{shazeer2017outrageouslylargeneuralnetworks}.

Let \(s_{l,e}\) denote the stored size of routed expert \(e\) in layer \(l\). The complete routed-expert footprint is
\begin{equation}\label{eq::footprint}
M_{\mathrm{routed}}
=
\sum_{l=1}^{L}
\sum_{e=1}^{N_l}
s_{l,e}.
\end{equation}
Thus, routed-expert storage is determined by the total expert set rather than by the number activated for each token. Routed expert weights consequently account for a large fraction of the model state that must be stored somewhere in the serving system.

\subsection{Expert Offloading on Heterogeneous Memory}
Compact AI appliances provide a heterogeneous memory hierarchy. GPU or NPU device memory offers high bandwidth for model execution but limited capacity, whereas host memory provides greater capacity at lower bandwidth. Expert offloading exploits the sparsity of MoE routing by keeping only a subset of experts resident in device memory and retaining the remaining experts in host memory \cite{xue2025moeinfinityefficientmoeinference,yi2025edgemoeempoweringsparselarge}.  When the router selects a non-resident expert, its weights must be transferred to the device before execution.

For a single device, expert residency must satisfy
\begin{equation}
\begin{aligned}
    M_{\mathrm{fixed}}
+ M_{\mathrm{KV}}(B,T)
+ M_{\mathrm{act}}(B,T)\\
+ M_{\mathrm{workspace}}
+ \sum_{l=1}^{L}\sum_{e=1}^{N_l} x_{l,e}s_{l,e}
\leq M_{\mathrm{device}},
\end{aligned}
\end{equation}
where \(x_{l,e}\in\{0,1\}\) indicates whether expert \(e\) in layer \(l\) is resident, \(B\) is the batch size, \(T\) is the context length, and \(M_{\mathrm{fixed}}\) denotes non-routed-expert weights and other persistent state.

Offloading improves memory feasibility, but a cache miss introduces host-to-device (H2D) communication. For an expert of size \(s_{l,e}\), its transfer time can be approximated as
\begin{equation}
    t^{\mathrm{H2D}}_{l,e}
\approx
\tau_{\mathrm{launch}}
+
\frac{s_{l,e}}{\beta_{\mathrm{H2D}}},
\end{equation}
where \(\beta_{\mathrm{H2D}}\) is the effective H2D bandwidth and \(\tau_{\mathrm{launch}}\) is the fixed transfer overhead. If the transfer begins only after experts are selected and cannot overlap with computation, its time lies on the inference critical path. Expert-offloading performance therefore depends on both the resident expert set and the ability to hide remaining transfers.

\subsection{Inference Phases and Memory Pressure}
Long-context Agent workloads illustrate the memory pressure of MoE inference: multi-step tasks accumulate conversation turns, retrieved evidence, tool outputs, and intermediate task state. AgentLongBench \cite{fang2026agentlongbenchcontrollablelongbenchmark} models such workloads with trajectories ranging from 32K to 4M tokens, spanning long-turn interactions and high-information-density tool responses.

MoE inference follows a prefill--decode workflow. Prefill processes the input context, typically in parallel, and constructs the KV cache; its latency is a major component of time to first token (TTFT). Decode reuses the cache to generate subsequent tokens iteratively, with time per output token (TPOT) characterizing its per-token latency \cite{298679,zhong2024distservedisaggregatingprefilldecoding}. In MoE models, prefill may access a broad expert set across many context tokens, whereas each decode step routes only the newly generated token or token group.

Longer trajectories expand the KV cache, leaving less device memory for resident experts. Context length and KV-cache representation therefore jointly determine the expert-residency budget.

\subsection{Observation and Motivation}\label{subsec::ham}
Memory constraints limit how many experts can remain resident on the device, while routing determines which experts are needed during inference. The resulting offloading cost depends on the interaction between expert residency and routing decisions, particularly when a selected expert is absent from device and its transfer cannot be hidden.

\begin{figure}[t]
    \centering
    \includegraphics[width=\linewidth]{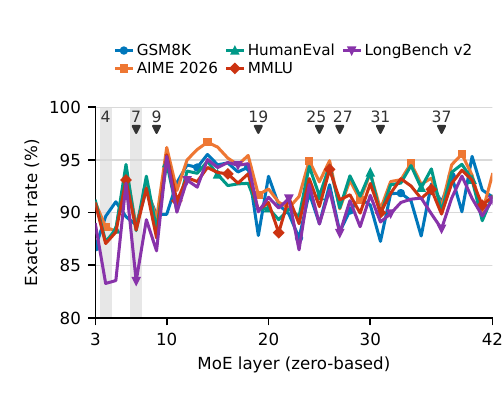}
    \caption{Exact selected-expert hit rate on DeepSeek-V4-Flash-W4A8 with 84 cached experts per layer. The y-axis spans 80--100\%. Gray bands mark layers 4 and 7, both bottom-ten in all five datasets. Triangles mark layers 7, 9, 19, 25, 27, 31, and 37, strict local troughs (below both neighbors) in all five.}
    \label{fig:five_dataset_layer_hits}
\end{figure}

\noindent\textbf{Equal Per-Layer Cache Sizes Lead to Uneven Hit Rates.} Figure~\ref{fig:five_dataset_layer_hits} compares five workloads with 84 resident experts per DeepSeek MoE layer, excluding zero-based layers 0--2, whose routing differs. Across layers 3--42, each workload's hit-rate range spans 7.64--12.07 percentage points. Layers 4 and 7 lie in the bottom ten in all five workloads; layers 7, 9, 19, 25, 27, 31, and 37 are strict local troughs in every trace (below both neighbors). Thus, a uniform per-layer quota yields recurrent coverage deficits at specific layers across the measured workloads. This motivates dynamic layer-wise allocation of a fixed expert-cache budget to improve coverage where additional capacity is most effective (Section~\ref{subsec:dynamicbuffer}).

\begin{figure}[t]
    \centering
    \includegraphics[width=\linewidth]{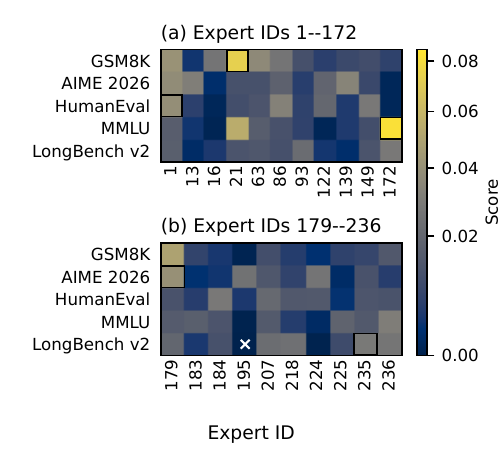}
    \caption{Normalized routing scores for DeepSeek-V4-Flash-W4A8 at layer 20. The union of the five Top-6 expert sets (21 IDs) is ordered by expert ID and split into two blocks. Black outlines mark each workload's top expert; the cross marks an omitted zero-score ID.}
    \label{fig:five_dataset_expert_rank}
\end{figure}

\noindent\textbf{Expert Routing May Exhibit Domain-Specific Bias.} Figure~\ref{fig:five_dataset_expert_rank} shows a representative comparison of five workloads at MoE layer 20. Their top-ranked expert IDs differ: 21 for GSM8K, 179 for AIME 2026, 1 for HumanEval, 172 for MMLU, and 235 for LongBench v2. Across the ten workload pairs at this layer, Spearman correlations over all 256 expert scores range from 0.546 to 0.784, indicating shared ranking structure. Yet Top-6 Jaccard overlap ranges from 0 to 0.200 (mean 0.116), and Top-32 overlap ranges from 0.306 to 0.455. Across the 40 MoE layers (3--42), the top-ranked expert is not identical across workloads in 37 layers, with a mean pairwise Top-6 Jaccard overlap of 0.184. This recurring divergence indicates domain-conditioned routing bias and motivates domain-informed cache initialization.

\begin{figure}[ht]
    \centering
    \includegraphics[width=\linewidth]{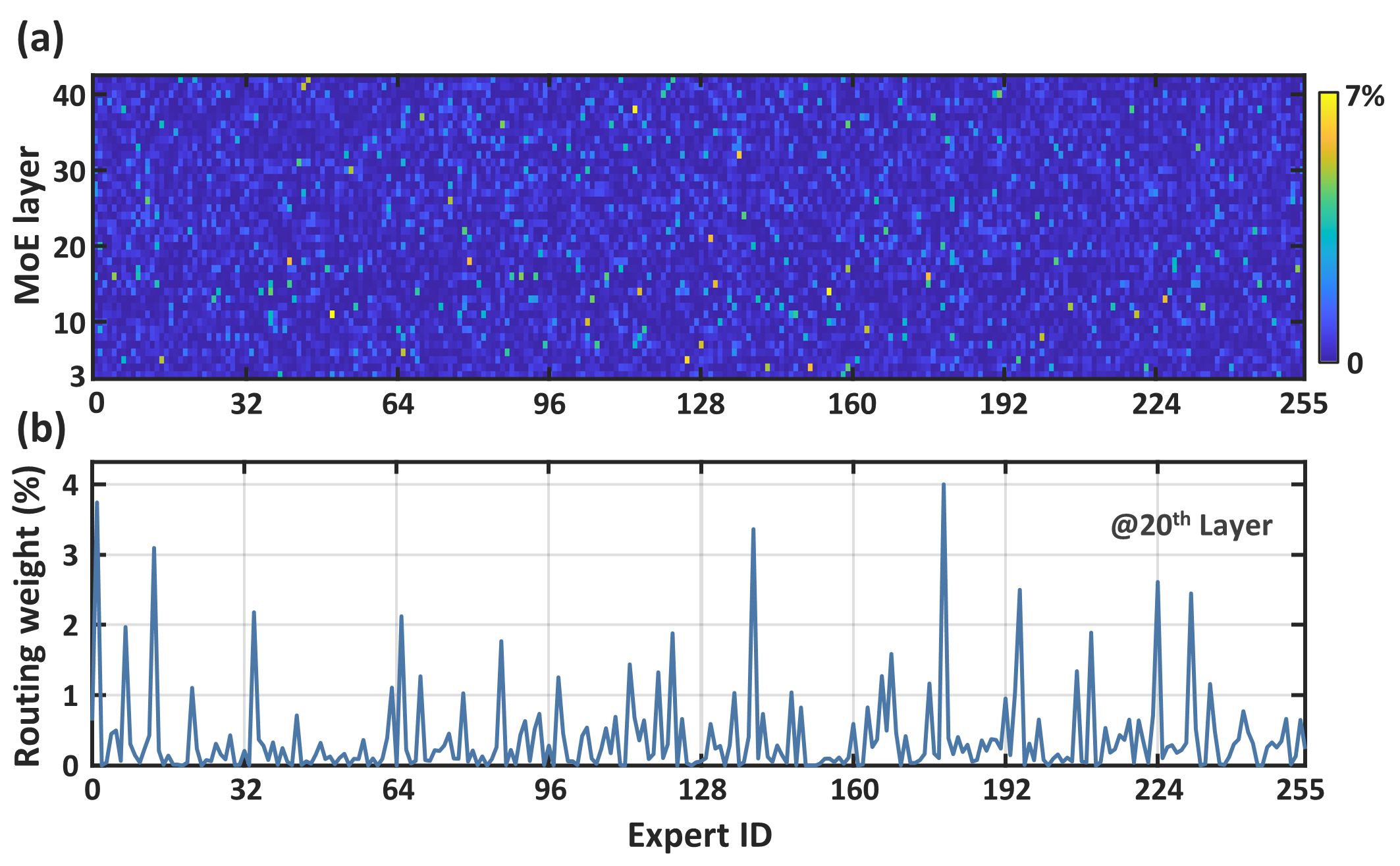}
    \caption{Illustrative DeepSeek-V4-Flash-W4A8 routing scores on AIME 2026. (a) Score distributions across MoE layers 3--42. (b) Scores by expert ID at layer 20, including the rank-6/rank-7 Top-6 boundary.}
    \label{fig:expert_weight_distribution}
\end{figure}

\noindent\textbf{Near-Boundary Scores Motivate Resident Expert Substitution.} We examine normalized routing scores across DeepSeek MoE layers 3--42 and zoom in on layer 20 (Figure~\ref{fig:expert_weight_distribution}). Panel (a) shows only a small group of high-scoring experts per layer; most experts occupy the low-score region. Panel (b) shows that the activation boundary can have a narrow score gap: the last activated expert (rank 6, ID 195) scores 2.5010\%, while the first non-activated expert (rank 7, ID 231) scores 2.4493\%, a difference of 0.0517 percentage points. Thus, expert selection need not imply a large score separation at the Top-6 cutoff. This motivates substituting a similar-scoring resident non-activated expert for an off-device low-score activated expert to bypass H2D transfer.

\section{MoE-CORE}
\begin{figure*}
    \centering
    \includegraphics[width=\linewidth]{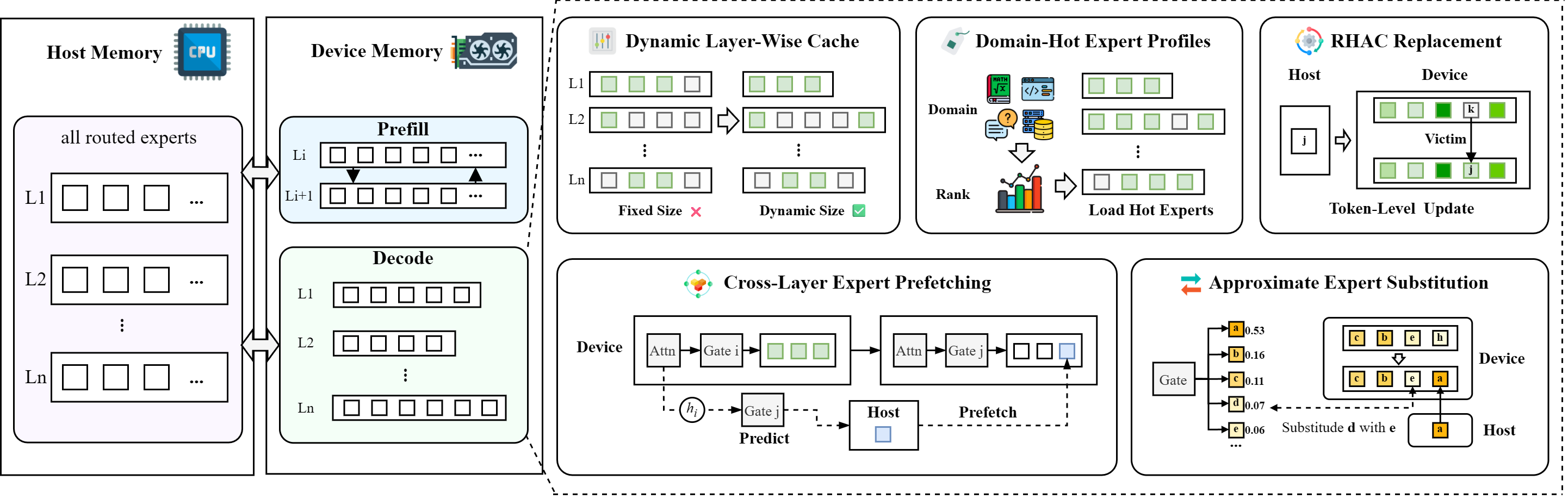}
    \caption{MoE-CORE's phase-specific expert-weight paths and decode-cache mechanisms. The prefill buffers stage complete MoE layers, whereas the decode region retains per-layer expert subsets and supports domain-based initialization, RHAC replacement, cross-layer prefetching, and optional substitution.}
    \label{fig:MoE-CORE}
\end{figure*}

MoE-CORE is a coordinated expert-weight offloading system for memory-constrained MoE inference on an AI appliance. Its design follows the phase asymmetry described in Section~\ref{subsec::ham}: prefill can collectively access a broad expert set, whereas each decode iteration exposes a sparse but latency-sensitive working set. Applying one offloading policy to both phases would either transfer unnecessary experts during decode or incur repeated expert-selection and loading overhead during long prefills. MoE-CORE therefore provides separate prefill and decode paths within a unified runtime; Figure~\ref{fig:MoE-CORE} shows the process.

The two paths use physically disjoint device-memory regions that remain allocated throughout inference. The prefill region contains two expert-weight buffers, each holding all routed experts from a single MoE layer. When the prefill token count reaches the cache-dependent threshold \(N_l^*\) defined in Section~\ref{subsec:prefill}, the buffers stage the expert weights of two consecutive layers and alternate as execution advances through the model. The decode region stores routed-expert weights across all MoE layers, organizing its capacity into per-layer caches with \(K_l\) resident expert slots assigned to layer \(l\). Both regions share the device-memory budget with the fixed model state, KV cache, activations, and runtime workspace.

During decode, MoE-CORE first establishes expert residency before attempting to hide communication. A layer-wise allocator determines \(K_l\). The serving application then specifies the workload domain through a request hint, and MoE-CORE fills the allocated slots with the corresponding hot experts identified by offline profiling. As decoding proceeds, a Routing-History-Aware Cache (RHAC) policy updates expert residency using recent activations, longer-term routing history, router scores, and access recency. These mechanisms reduce cache misses but cannot eliminate them. MoE-CORE then adopts a cross-layer prefetcher that predicts the next layer's routed experts and asynchronously transfers predicted experts during the intervening computation. If a selected expert remains unavailable when required, the exact path loads it from host memory. When optional approximate expert substitution is enabled, an eligible low-score miss may instead use a score-similar, unselected expert already resident on the device; high-score misses remain on the exact path.

Thus, the decode path combines miss reduction with communication overlap before considering approximate execution. Layer-wise allocation, domain-profile initialization, RHAC, and cross-layer prefetching preserve the routed expert computation; only optional expert substitution changes the selected expert set.

\subsection{Dynamic Layer-Wise Expert-Cache Allocation}\label{subsec:dynamicbuffer}
The decode expert region has a fixed memory capacity, but that capacity need not be divided equally among MoE layers. Let \(M_{\mathrm{dec}}\) be the memory assigned to the decode expert cache, \(K_l\) the number of expert slots allocated to layer \(l\), and \(S_l\) the size of one expert in that layer. The allocation must satisfy $\sum_{l=1}^{L} K_l S_l \leq M_{\mathrm{dec}}$.

A uniform design assigns the same number of slots to every layer. Section~\ref{subsec::ham} identifies low-hit layers that recur across the measured workloads, making them candidates for additional slots. A low hit rate does not by itself establish that more capacity will help, so MoE-CORE profiles each layer's response to cache-size changes before redistributing the decode expert cache.

The allocator first replays representative routing traces and measures the layer-wise hit-rate curve \(\widehat H_l(k)\) for a small set of candidate cache sizes around the uniform allocation. The replacement and initialization policies are held constant during this profiling so that the measured difference reflects only the number of slots. Starting from the uniform allocation, the allocator moves slots in increments of \(\Delta K\). For the current allocation \(K_l\), it estimates the loss caused by removing \(\Delta K\) slots and the gain obtained by adding them as

\begin{equation}
\begin{aligned}
    D_l^{-}&=\widehat H_l(K_l)-\widehat H_l(K_l-\Delta K),
    \\
    G_l^{+}&=\widehat H_l(K_l+\Delta K)-\widehat H_l(K_l).
\end{aligned}
\end{equation}

At each allocation round, layers whose hit rates remain above the layer-wise median and whose removal loss is no greater than \(\epsilon_{\mathrm{drop}}\) form the donor set. Layers whose hit rates fall at least \(\tau_{\mathrm{low}}\) below the median form the receiver set. MoE-CORE selects the donor \(d\) with the smallest \(D_d^{-}\) and the receiver \(r\) with the largest \(G_r^{+}\). If \(G_r^{+}>D_d^{-}\) and the resulting allocation satisfies the memory constraint, it updates

\begin{equation}
K_d \leftarrow K_d-\Delta K,
\qquad
K_r \leftarrow K_r+\Delta K.
\end{equation}

The donor and receiver sets are recomputed after every transfer. Reallocation stops when no receiver remains below the low-hit threshold, no donor can release slots within the permitted drop, or no feasible transfer offers positive estimated gain. The resulting allocation \(\{K_l\}_{l=1}^{L}\) remains within the decode-cache budget while shifting capacity from locally insensitive high-hit layers to cache-sensitive low-hit layers.

Profiling and allocation run outside the per-token inference path and are repeated when the model, deployment workload, or available decode-cache capacity changes. The allocator determines only how many expert slots each layer receives. The selected workload-domain profile determines which experts initially fill those slots, and RHAC subsequently updates their contents during decoding. Thus, layer-wise allocation changes neither the router decisions nor the model computation.

\subsection{Offline Domain Hot-Expert Profiles}
Agent workloads span distinct application domains, including software engineering, mathematical reasoning, and domain-specific tool use \cite{jimenez2024swe,ICLR2024_d3cf1559,yao2024tau}. The domain-dependent expert rankings observed in Section~\ref{subsec::ham} motivate domain-conditioned cache initialization. For each supported domain, MoE-CORE constructs an offline hot-expert profile. At request admission, the serving scheduler supplies a domain identifier \(d\), and the corresponding profile determines which experts initially occupy the allocated cache slots.

We construct these profiles for a fixed set of commonly served domains using mathematical-reasoning, code-generation, and QA datasets. For domain \(d\), let \(\mathcal{T}_d\) denote all tokens in its profiling dataset, and let \(\mathcal{E}_l=\{1,\ldots,N_l\}\) be the complete expert set at layer \(l\). For every token \(t\in\mathcal{T}_d\), the profiler records the selection score \(r_{l,t,e}\), the \(e\)-th component of \(\mathbf r_{l,t}\), for every expert \(e\in\mathcal{E}_l\) before sparse expert selection. The mean selection score of expert \(e\) is
\begin{equation}
    \mu_{d,l,e}=
\frac{1}{|\mathcal{T}_d|}
\sum_{t\in\mathcal{T}_d}r_{l,t,e}.
\end{equation}
The vector \(\boldsymbol{\mu}_{d,l}=(\mu_{d,l,1},\ldots,\mu_{d,l,N_l})\) summarizes the domain-conditioned selection preference over all experts in layer \(l\). For every domain--layer pair, the profiler stores a permutation \(\pi_{d,l}\) satisfying
\begin{equation}
    \mu_{d,l,\pi_{d,l}(1)}
\geq
\mu_{d,l,\pi_{d,l}(2)}
\geq \cdots \geq
\mu_{d,l,\pi_{d,l}(N_l)}.
\end{equation}
Once the layer-wise allocator assigns \(K_l\) cache slots to layer \(l\), MoE-CORE initializes the resident set with the \(K_l\) experts having the highest mean selection scores:
\begin{equation}
    \mathcal{R}^{init}_l(d)
=
\left\{
\pi_{d,l}(1),\ldots,\pi_{d,l}(K_l)
\right\}.
\end{equation}
This construction separates cache capacity from cache contents: the layer-wise allocator determines how many experts layer \(l\) can retain, while the selected domain profile determines which experts initially occupy those slots.

\subsection{Routing-History-Aware Expert Cache}
The offline layer-wise allocation and domain profiles cannot capture routing changes within a session. MoE-CORE therefore manages each layer's resident experts with RHAC. Like LRU, RHAC selects eviction victims, but it bases the decision on both activation history and the router scores already produced during MoE inference.

RHAC is updated per token. Let \(i=1,2,\ldots\) denote the update order, let \(t_i\) be the token processed at update \(i\), and let \(W\in\mathbb{N}^{+}\) denote the fixed history-window length in tokens. RHAC maintains four signals for every expert \(e\) at layer \(l\): recent selection frequency $f$, long-term selection tendency $\bar f$, the current router score $q$, and age $a$. Before the first generated token, the recent-frequency window is empty and the two recurrent states are initialized as \(\bar f^{(0)}_{l,e}=a^{(0)}_{l,e}=0\).

First, the recent selection frequency counts how many of the latest \(W\) tokens select expert \(e\):
\begin{equation}
f^{(i)}_{l,e}
=
\sum_{j=\max(1,i-W+1)}^{i}
\mathbb{I}\!\left[e\in\mathcal{S}_{l,t_j}\right].
\end{equation}
Thus, \(f^{(i)}_{l,e}\) ranges from zero to \(\min(i,W)\). Second, RHAC captures a longer-term selection tendency with an exponential moving average (EMA) \cite{brown1956exponential},
\begin{equation}
\bar f^{(i)}_{l,e}
=
\rho_{\mathrm{sel}}\bar f^{(i-1)}_{l,e}
+(1-\rho_{\mathrm{sel}})
\mathbb{I}\!\left[e\in\mathcal{S}_{l,t_i}\right].
\end{equation}
Here, \(0\leq\rho_{\mathrm{sel}}<1\) is the selection-history decay factor. Values closer to one retain earlier selections for longer, whereas values closer to zero emphasize the current token. Third, the current router scores provide information about every expert. Let $q_{l,t,e}=r_{l,t,e}$ denote the expert-selection score assigned to expert \(e\) for token \(t\). RHAC uses the current value \(q_{l,t_i,e}\) directly. Finally, age records how many generated tokens have elapsed since the expert was last selected:
\begin{equation}
    a^{(i)}_{l,e}
=
\begin{cases}
0, & e\in\mathcal{S}_{l,t_i},\\
a^{(i-1)}_{l,e}+1, & e\notin\mathcal{S}_{l,t_i}.
\end{cases}
\end{equation}
Because the four signals have different scales, each is normalized across all \(N_l\) experts in the same layer before combination; a tilde denotes the resulting layer-wise normalized value. RHAC assigns expert \(e\) the residency-priority score
\begin{equation}
\eta^{(i)}_{l,e}
=
\lambda_f\widetilde f^{(i)}_{l,e}
+\lambda_m\widetilde{\bar f}^{(i)}_{l,e}
+\lambda_q\widetilde q_{l,t_i,e}
-\lambda_a\widetilde a^{(i)}_{l,e},
\end{equation}
where the four nonnegative coefficients \(\lambda_f\), \(\lambda_m\), \(\lambda_q\), and \(\lambda_a\) control the contributions of recent frequency, the selection EMA, the current router score, and age, respectively. Recent and sustained selections raise the priority, a high router score protects an expert that is relevant to the present token, and increasing age lowers the priority of stale experts.

When a routed miss must enter a full cache, RHAC selects the coldest evictable resident expert,
\begin{equation}
    e^{\mathrm{victim}}_{l,i}
=
\arg\min_{e\in\mathcal{R}_l\setminus\mathcal{P}_{l,i}}
\eta^{(i)}_{l,e},
\end{equation}
where \(\mathcal{R}_l\) is the current resident set and \(\mathcal{P}_{l,i}\) contains experts that are executing, loading, or otherwise protected from eviction while token \(t_i\) is processed. After the victim is released and the requested weights arrive, the new expert occupies the slot. Therefore, RHAC remains free to adapt the cache to the observed session.

\subsection{Cross-Layer Expert Prefetching}
\begin{figure}
    \centering
    \includegraphics[width=\linewidth]{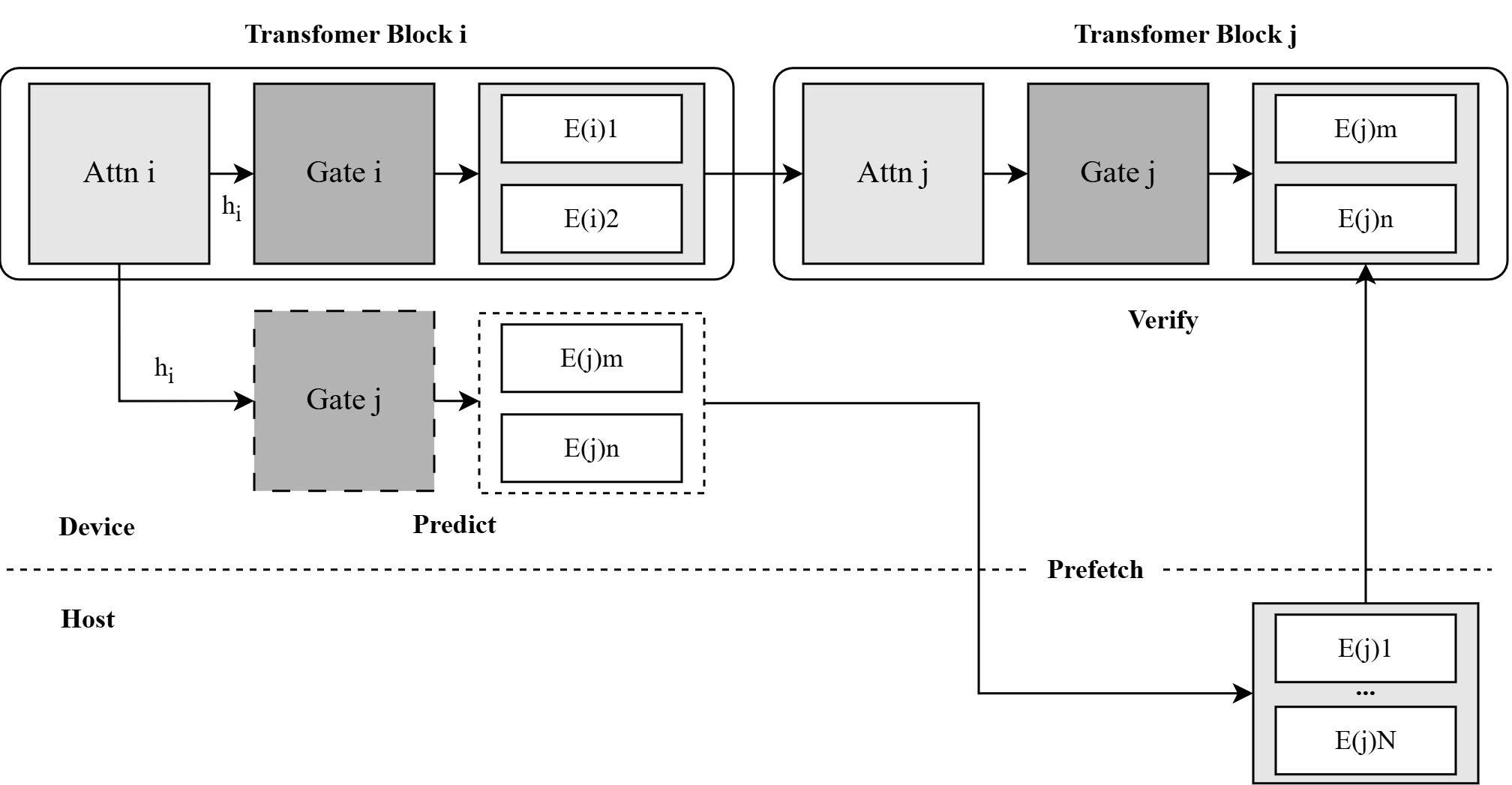}
    \caption{Cross-layer expert prefetching. The next layer's router receives the current layer's hidden state to rank candidate experts before its ordinary input becomes available; predicted experts are transferred during intervening computation.}
    \label{fig:prefecth}
\end{figure}

Even RHAC cannot eliminate every decode expert miss. Loading the missed expert only after the true router finishes places the complete H2D transfer on the critical path. MoE-CORE predicts the next layer's expert ranking early and transfers a bounded number of likely experts while the model continues computing, as illustrated in Figure~\ref{fig:prefecth}.

Section~\ref{subsec::moe} defines \(h_{l,t}\) as the hidden state used for routing token \(t\) at layer \(l\). Before the normal router input \(h_{l+1,t}\) becomes available, MoE-CORE evaluates the original router of layer \(l+1\) on \(h_{l,t}\):
\begin{equation}
\widehat{\mathbf r}_{l+1,t}
=
R_{l+1}(h_{l,t}).
\end{equation}
Here, \(\widehat{\mathbf r}_{l+1,t}\) is the predicted expert-score vector for layer \(l+1\), and MoE-CORE prefetches experts according to its ranking. This prediction is motivated by cross-layer routing consistency: routing decisions in adjacent MoE layers can remain correlated, allowing \(R_{l+1}(h_{l,t})\) to approximate the ranking later produced by \(R_{l+1}(h_{l+1,t})=\mathbf r_{l+1,t}\) \cite{Fate2026Fang,Zhong_2024}.

The prefetch count is determined by the computation window available for hiding communication. Let
\begin{equation}
T^{\mathrm{overlap}}_l
=
T^{\mathrm{remaining}}_l
+T^{\mathrm{attn}}_{l+1}
\end{equation}
be the interval from the early prediction until the true layer-\(l+1\) gate completes. It contains the remaining computation of layer \(l\) and the attention computation of layer \(l+1\). If the effective H2D bandwidth is \(\beta_{\mathrm{H2D}}\), the transfer launch overhead is \(\tau_{\mathrm{H2D}}\), and each layer-\(l+1\) expert occupies \(s_{l+1}\) bytes, an upper bound on the number of complete expert transfers hidden by this interval is
\begin{equation}
n^{\mathrm{hide}}_{l+1}
=
\left\lfloor
\frac{
\max(0,T^{\mathrm{overlap}}_l-\tau_{\mathrm{H2D}})
\beta_{\mathrm{H2D}}
}{s_{l+1}}
\right\rfloor.
\end{equation}
Let \(n^{\mathrm{cand}}_{l+1}\) be the number of predicted experts that are neither resident nor already loading, and let \(n^{\mathrm{slot}}_{l+1}\) be the number of slots the runtime can make available without evicting an in-use expert. MoE-CORE prefetches
\begin{equation}\label{eq::h2dexperts}
    n^{\mathrm{pf}}_{l+1}
=
\min\!\left(
n^{\mathrm{hide}}_{l+1},
n^{\mathrm{cand}}_{l+1},
n^{\mathrm{slot}}_{l+1}
\right)
\end{equation}
highest-ranked predicted experts. Their H2D copies are issued asynchronously. When the true router completes, an exact expert that has arrived can execute immediately; an incorrect prefetch is discarded or retained according to RHAC, and an expert that remains unavailable becomes a residual miss. This bound limits the scheduled transfers to the estimated overlap window; inaccurate predictions may nevertheless consume bandwidth and cache capacity.

\subsection{Approximate Expert Substitution}
\begin{figure}
    \centering
    \includegraphics[width=\linewidth]{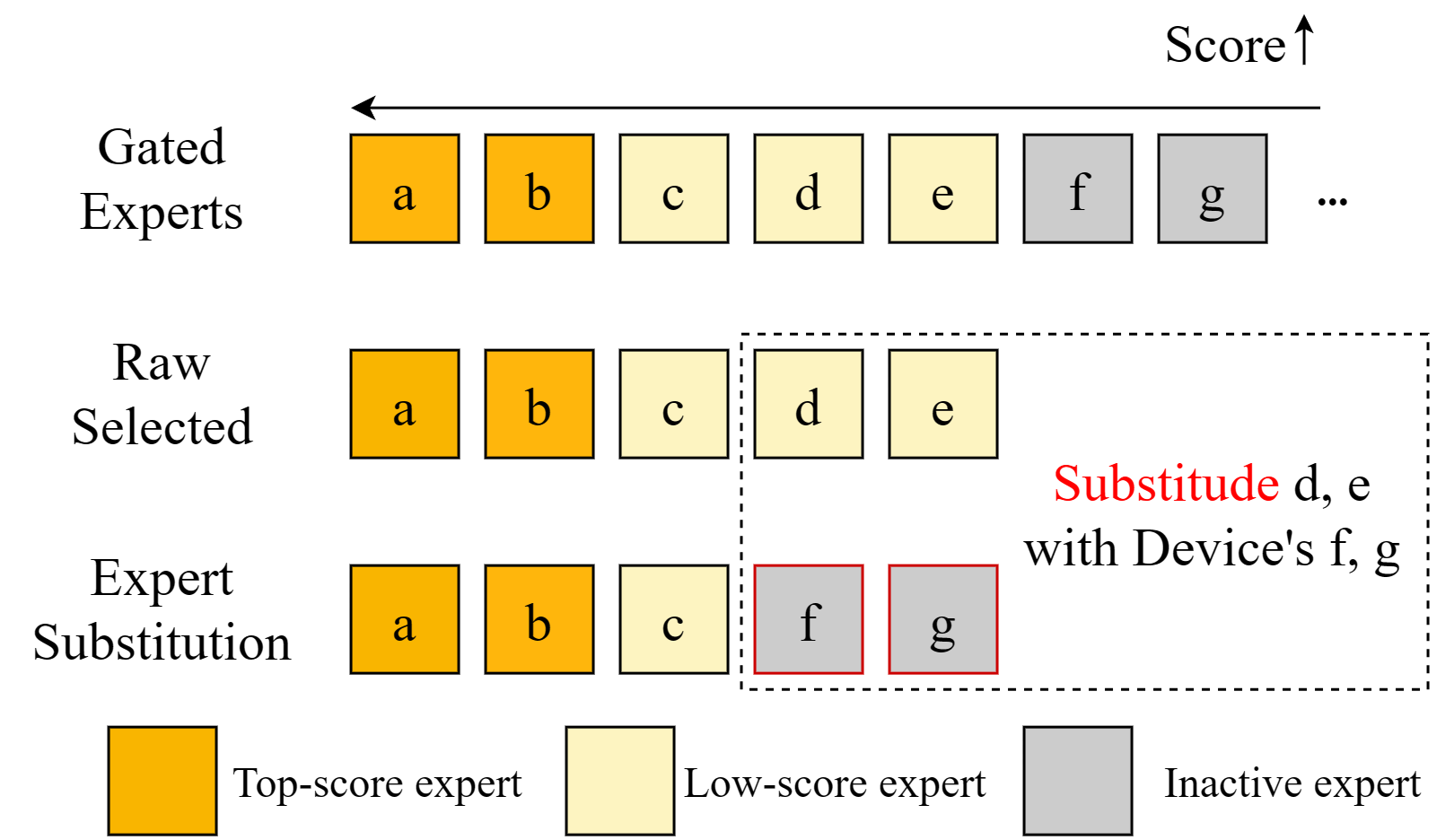}
    \caption{Illustrative expert substitution near the Top-\(k\) boundary. Selected experts d and e are off-device, while lower-ranked unselected experts f and g are resident. Replacement is considered only when their routing scores satisfy the proximity criterion.}
    \label{fig:expertsub}
\end{figure}

Cross-layer prefetching reduces exposed communication but cannot guarantee that every selected expert is available. Synchronously loading every residual miss preserves exact execution but can introduce a sudden latency spike for the affected token. Thus, MoE-CORE provides an optional approximate expert substitution (AES) path based on the score distribution around the Top-\(k\) routing boundary. The mechanism uses score proximity to restrict substitution to low-score routing decisions near the selection boundary. Figure~\ref{fig:expertsub} illustrates an example of the substitution process.

For token \(t\) at layer \(l\), let \(r_{l,t,e}\) denote the score of expert \(e\) in the router-score vector \(\mathbf r_{l,t}\). Order all \(N_l\) experts such that
\begin{equation}
    r_{l,t,(1)}\ge r_{l,t,(2)}\ge\cdots\ge r_{l,t,(N_l)}.
\end{equation}
The selected set is \(\mathcal{S}_{l,t}=\{(1),\ldots,(k_l)\}\), and \(b_{l,t}=r_{l,t,(k_l+1)}\) is the score immediately below the Top-\(k\) boundary. Given a threshold hyperparameter \(0\le\alpha<1\), MoE-CORE partitions the selected experts into an exact-only high-score set
\begin{equation}
    \mathcal{E}^{\mathrm{high}}_{l,t}
=
\left\{
e\in\mathcal{S}_{l,t}:
r_{l,t,e}>(1+\alpha)b_{l,t}
\right\},
\end{equation}
and a replaceable low-score set
\begin{equation}
    \mathcal{E}^{\mathrm{low}}_{l,t}
=
\left\{
e\in\mathcal{S}_{l,t}:
b_{l,t}\le r_{l,t,e}\le(1+\alpha)b_{l,t}
\right\}.
\end{equation}

Every cache miss in \(\mathcal{E}^{\mathrm{high}}_{l,t}\) follows the exact H2D path. Among the low-score selected experts, let
\begin{equation}
    \mathcal{U}_{l,t}
=
\mathcal{E}^{\mathrm{low}}_{l,t}\setminus\mathcal{R}_l
=
\{u_1,\ldots,u_U\}
\end{equation}
be the residual misses that remain unavailable after prefetching. A substitution candidate must be unselected, already resident, and sufficiently close to the lower side of the routing boundary:
\begin{equation}
\begin{aligned}
\mathcal{C}_{l,t}&=
\left\{
c\in\mathcal{R}_l\setminus\mathcal{S}_{l,t}:
(1-\alpha)b_{l,t}\le r_{l,t,c}\le b_{l,t}
\right\}\\
& =
\{c_1,\ldots,c_C\}.
\end{aligned}
\end{equation}
MoE-CORE sorts both sets by decreasing router score and performs a one-to-one matching:
\begin{equation}
    \begin{aligned}
r_{l,t,u_1}&\ge\cdots\ge r_{l,t,u_U},\\
r_{l,t,c_1}&\ge\cdots\ge r_{l,t,c_C},\\
u_p&\leftarrow c_p,
\qquad p=1,\ldots,\min(U,C).
\end{aligned}
\end{equation}
Thus, the highest-scoring resident candidate replaces the highest-ranked eligible miss, the next candidate replaces the next miss, and no candidate is reused. Because all candidates rank below all selected experts, this ordered matching assigns each miss the closest remaining candidate in routing rank. Any high-score miss and any unmatched low-score miss remain on the exact H2D path.

Let \(P=\min(U,C)\) and \(\mathcal{U}^{\mathrm{sub}}_{l,t}=\{u_1,\ldots,u_P\}\) denote the matched misses. For each pair \((u_p,c_p)\), the candidate expert \(c_p\) replaces only the expert function, while retaining the missed expert's original routing coefficient \(\alpha_{l,t,u_p}\). The routed MoE output becomes
\begin{equation}\label{eq::moeoutputnew}
\begin{aligned}
    m^{\mathrm{sub}}_{l,t}
=
&\sum_{e\in\mathcal{S}_{l,t}\setminus\mathcal{U}^{\mathrm{sub}}_{l,t}}
\alpha_{l,t,e}E_{l,e}(h_{l,t})
+\\
&\sum_{p=1}^{P}
\alpha_{l,t,u_p}E_{l,c_p}(h_{l,t}).
\end{aligned}
\end{equation}
The candidate's own router score is used only for matching and does not replace the missed expert's routing coefficient. Eq.~\ref{eq::moeoutputnew} changes only the matched near-boundary expert computations and removes their synchronous H2D transfers from the token's critical path. The threshold \(\alpha\) limits the score range eligible for substitution. Score proximity is a selection heuristic and does not establish equivalence between the original and substitute expert outputs.

\subsection{Layer-Wise Expert Staging for Prefill}\label{subsec:prefill}
Prefill processes many tokens together, so the union of their routing decisions can cover most experts in a layer. Motivated by the observation that complete expert loading can become preferable to selective loading as the token count increases \cite{wang2025acceleratingmixtureofexpertsinferencehiding}, MoE-CORE uses the initial per-layer resident-expert count to set the staging threshold. For layer \(l\), let \(K_l^{(0)}\) denote the initially configured number of cached experts and let \(k_l\) denote the routing fanout (\texttt{topk}). The threshold is computed as
\begin{equation}
    N_l^* = \left\lfloor \frac{K_l^{(0)}}{k_l} \right\rfloor.
\end{equation}
Thus, \(N_l^*\) varies with the configured per-layer cache capacity and routing fanout, rather than being a fixed token count. Layers for which the prefill has at least \(N_l^*\) tokens use complete layer-wise staging; smaller prefills retain selective per-layer residency.

The physically reserved prefill region contains two buffers, \(B_0\) and \(B_1\), each dedicated to the routed-expert weights of one MoE layer. Each buffer is sized for the routed-expert footprint of the largest MoE layer, and the prefill region reserves twice this capacity.

For the complete-staging path, MoE-CORE loads the first two MoE layers into \(B_0\) and \(B_1\) during warm-up. After layer \(l\) finishes using one buffer, that buffer becomes available for layer \(l+2\). While layer \(l+1\) computes from the other buffer, the transfer engine loads all experts of layer \(l+2\) into the released buffer. The two buffers alternate until the final MoE layer.

Ignoring warm-up and drain costs, the steady-state latency around layer \(l\) is approximated by
\begin{equation}
    T^{\mathrm{steady}}_l
\approx
\max\!\left(
T^{\mathrm{compute}}_l,
T^{\mathrm{H2D}}_{l+1}
\right)
+T^{\mathrm{sync}}_l.
\end{equation}
If the current layer's computation is at least as long as the next layer's transfer, the transfer can be fully hidden after pipeline warm-up. Otherwise, only their difference remains exposed. The initial loading of the first two layers, the final pipeline drain, and all synchronization costs remain part of TTFT.

\section{Evaluation}
We evaluate end-to-end inference performance, the contributions of the expert-cache components, and compatibility with multi-token prediction (MTP). The main comparisons and cache-component ablations use exact expert execution; AES and MTP are evaluated separately. Unless otherwise specified, Deepseek is the default model.

\subsection{Implementation Settings}
\noindent\textbf{Datasets and Models.} We use GSM8K \cite{cobbe2021trainingverifierssolvemath}, AIME 2026 \cite{dekoninck2026matharena}, HumanEval \cite{chen2021evaluatinglargelanguagemodels}, MMLU \cite{hendrycks2021measuringmassivemultitasklanguage}, and LongBench v2 \cite{bai2025longbenchv2deeperunderstanding}. They cover mathematical reasoning, code generation, knowledge, and long-context tasks. The evaluated models are DeepSeek-V4-Flash-W4A8 \cite{deepseekai2026deepseekv4} and GLM-5.2-W4A8C8 \cite{glm5team2026glm5vibecodingagentic}.

\noindent\textbf{Platforms.} Both models are evaluated on an Ascend 950PR NPU platform with 128 GB of physical NPU memory and a 150-GB host-memory limit. Inference is subject to an 84-GB NPU-memory cap.

\noindent\textbf{Baselines.} We compare MoE-CORE with two expert-offloading baselines:
\begin{itemize}
    \item \textbf{Load on Demand (LoD).} The baseline allocates static per-layer expert caches only during decode and synchronously loads a selected expert on a cache miss. 
    \item \textbf{vLLM Prefetch.} We use the layer-wise prefetch strategy in vLLM \cite{vllm_prefetch_offloader}. The DeepSeek configuration uses two resident layer buffers and two rotating layer buffers; the GLM configuration uses one resident layer and five offloaded layers.
\end{itemize}

\noindent\textbf{Request Protocol.} All configurations in Table~\ref{tab:main_performance} use batch size 1. LoD and MoE-CORE allow up to 1K output tokens, whereas vLLM Prefetch is capped at 128 output tokens to bound the runtime of its substantially slower decoding. DeepSeek and GLM use 200 requests per dataset.

\noindent\textbf{Metrics.} We report time to first token (TTFT), time per output token (TPOT), end-to-end latency (E2EL), and output throughput in tokens per second (TPS). For a response containing \(n>1\) output tokens, TPOT is the generation time after the first token divided by \(n-1\). TTFT-P90 and TPOT-P90 are the reported request-level 90th percentiles. Because the output caps differ, E2EL and TPS comparisons are restricted to LoD and MoE-CORE. Exact cache-hit rate measures the availability of the originally selected experts, including prefetched experts ready before execution. For AES-enabled results, effective expert availability additionally counts selected invocations served by resident substitutes. Means computed from the provided layer-rate arrays weight the recorded layers equally.

\subsection{Implementation Details}
MoE-CORE reserves separate expert-weight regions for prefill staging and decode caches. Of the 84-GB NPU-memory cap, 95\% is budgeted for model inference, including model weights, expert caches, KV cache, activations, and inference workspace; the remaining 5\% is reserved for other necessary system allocations. The maximum configured prompt-plus-output length is 368K tokens. In the DeepSeek configuration, the uniform decode allocation uses 84 expert slots per layer. The evaluated nonuniform allocation uses 78 slots for most layers and 120 slots for layers 7, 9, 19, 31, and 37. GLM's larger expert weights limit its single-NPU decode cache to 28 expert slots per layer. Domain profiles initialize the resident experts, and RHAC updates residency once per generated token. Cross-layer prefetching fetches one predicted expert per layer by default. The substitution threshold is \(\alpha=0.1\). AES is disabled in the main comparison and for the cache-component ablations.

\subsection{Effectiveness Evaluation}
Table~\ref{tab:main_performance} reports the measurements on each model. The MoE-CORE runs use exact expert execution.
% Source: MOEData workbook, Sheet1. LoD: rows 14--18 and 20--24.
% Prefetch: rows 3--7 and 43--47. MoE-CORE: rows 33--37 and 38--42.
% DeepSeek full-system data use the author-selected "second test" block.
\begin{table*}[t]
\centering
\caption{Inference performance at batch size 1 with output caps of 1K tokens for LoD and MoE-CORE and 128 tokens for vLLM Prefetch. Both models run on Ascend 950PR. LoD uses \texttt{num\_device\_layers=1}. AES is disabled in both MoE-CORE configurations, and MTP is disabled in the DeepSeek configuration. Lower latency and higher TPS are better; the best measured value for each dataset, model, and metric is underlined.}
\label{tab:main_performance}
\scriptsize
\setlength{\tabcolsep}{1.8pt}
\resizebox{\textwidth}{!}{%
\begin{tabular}{@{}ll*{12}{r}@{}}
\toprule
\textbf{Dataset} & \textbf{System}
& \multicolumn{6}{c}{\textbf{DeepSeek-V4-Flash-W4A8}}
& \multicolumn{6}{c}{\textbf{GLM-5.2-W4A8C8}} \\
\cmidrule(lr){3-8}\cmidrule(lr){9-14}
& & \multicolumn{2}{c}{\textbf{TTFT (ms)}}
& \multicolumn{2}{c}{\textbf{TPOT (ms)}}
& \textbf{E2EL (s)} & \textbf{TPS}
& \multicolumn{2}{c}{\textbf{TTFT (ms)}}
& \multicolumn{2}{c}{\textbf{TPOT (ms)}}
& \textbf{E2EL (s)} & \textbf{TPS} \\
\cmidrule(lr){3-4}\cmidrule(lr){5-6}
\cmidrule(lr){9-10}\cmidrule(lr){11-12}
& & \textbf{Mean} & \textbf{P90} & \textbf{Mean} & \textbf{P90} & &
& \textbf{Mean} & \textbf{P90} & \textbf{Mean} & \textbf{P90} & & \\
\midrule
GSM8K & LoD
& 3038.4 & 3094.8 & 66.5 & 71.9 & 37.84 & 13.57
& 8760.3 & 8920.8 & 233.9 & 239.2 & 180.84 & 4.07 \\
& vLLM Prefetch
& \underline{1175.1} & \underline{1177.7} & 1268.9 & 1268.9 & 162.00 & 0.78
& \underline{5662.7} & \underline{5664.3} & 5941.5 & 5941.6 & 760.24 & 0.17 \\
& MoE-CORE
& 3082.4 & 3124.4 & \underline{38.0} & \underline{39.2} & \underline{21.10} & \underline{22.35}
& 8950.0 & 9087.0 & \underline{210.3} & \underline{217.1} & \underline{152.94} & \underline{4.47} \\
\addlinespace
AIME 2026 & LoD
& 2948.4 & 2984.5 & 62.9 & 64.2 & 67.34 & 15.21
& 8860.2 & 8713.3 & 244.8 & 249.1 & 259.27 & 3.95 \\
& vLLM Prefetch
& \underline{1177.6} & \underline{1179.3} & 1268.9 & 1269.1 & 162.00 & 0.79
& \underline{5662.1} & \underline{5664.5} & 5941.5 & 5941.6 & 760.24 & 0.17 \\
& MoE-CORE
& 3026.2 & 3069.2 & \underline{42.1} & \underline{46.5} & \underline{46.09} & \underline{22.22}
& 9028.6 & 8843.4 & \underline{220.5} & \underline{226.8} & \underline{234.65} & \underline{4.36} \\
\addlinespace
HumanEval & LoD
& 2988.4 & 3025.2 & 62.1 & 63.4 & 36.38 & 14.78
& 8408.8 & 8910.0 & 235.6 & 241.4 & 182.05 & 4.05 \\
& vLLM Prefetch
& \underline{1178.6} & \underline{1181.6} & 1269.0 & 1269.0 & 162.00 & 0.79
& \underline{5663.2} & \underline{5664.7} & 5941.7 & 5941.7 & 760.25 & 0.17 \\
& MoE-CORE
& 3112.2 & 3153.0 & \underline{44.8} & \underline{50.0} & \underline{26.79} & \underline{19.17}
& 8515.2 & 9035.5 & \underline{212.9} & \underline{219.5} & \underline{175.80} & \underline{4.50} \\
\addlinespace
MMLU & LoD
& 2959.7 & 3019.2 & 64.1 & 65.4 & 61.20 & 14.81
& 8889.3 & 8910.8 & 229.1 & 234.9 & \underline{86.19} & 3.93 \\
& vLLM Prefetch
& \underline{1180.7} & \underline{1184.5} & 1269.0 & 1269.0 & 162.00 & 0.79
& \underline{5932.3} & \underline{5935.2} & 5941.8 & 5941.8 & 760.54 & 0.17 \\
& MoE-CORE
& 3044.4 & 3106.4 & \underline{41.5} & \underline{42.4} & \underline{43.90} & \underline{22.39}
& 9032.5 & 9118.5 & \underline{206.6} & \underline{209.6} & 91.43 & \underline{4.38} \\
\addlinespace
LongBench v2 & LoD
& 14056.4 & 31351.1 & 70.6 & 75.4 & 74.08 & 11.40
& 158894.0 & 290566.0 & 242.7 & 249.2 & 377.77 & 2.39 \\
& vLLM Prefetch
& \underline{9749.7} & \underline{19006.3} & 1269.1 & 1269.8 & 89.00 & 0.71
& 110850.0 & 202840.2 & 5941.8 & 5942.0 & 865.46 & 0.15 \\
& MoE-CORE
& 11879.0 & 24029.0 & \underline{40.3} & \underline{45.0} & \underline{47.08} & \underline{18.77}
& \underline{101463.4} & \underline{195094.5} & \underline{210.6} & \underline{217.2} & \underline{278.90} & \underline{3.01} \\
\bottomrule
\end{tabular}%
}
\end{table*}

On DeepSeek, MoE-CORE reduces mean TPOT from 62.1--70.6 ms to 38.0--44.8 ms. Relative to LoD, the reductions are 42.9\% on GSM8K, 33.1\% on AIME 2026, 27.9\% on HumanEval, 35.3\% on MMLU, and 42.9\% on LongBench v2, corresponding to \(1.39\text{--}1.75\times\) speedups. E2EL improves by \(1.36\text{--}1.79\times\) across the five workloads. TTFT increases by 1.4--4.1\% on the first four workloads; on LongBench v2 it falls from 14.06 to 11.88 s, while E2EL falls from 74.08 to 47.08 s.

On GLM, MoE-CORE records lower mean TPOT than LoD on all five workloads, at 206.6--220.5 versus 229.1--244.8 ms. On LongBench v2, its TPOT is 210.6 versus 242.7 ms and its E2EL is 278.90 versus 377.77 s. E2EL is lower on four of the five workloads; MMLU is the exception, increasing from 86.19 to 91.43 s. Across both models, MoE-CORE records lower TPOT than LoD on all ten model--workload pairs and lower E2EL on nine, demonstrating consistent decode-latency gains and favorable end-to-end performance across the evaluated workloads.

Under the stated output caps, the proposed method achieves lower TPOT than vLLM Prefetch on every dataset: 38.0--44.8 versus 1268.9--1269.1 ms on DeepSeek, and 206.6--220.5 versus 5941.5--5941.8 ms on GLM. vLLM Prefetch, in turn, records lower TTFT on all DeepSeek datasets and four GLM datasets. This trade-off reflects the different expert-staging choices. vLLM Prefetch stages complete expert layers and uses four layer buffers in the DeepSeek configuration, favoring prefill relative to MoE-CORE's two alternating full-layer buffers. During decode, however, the computation of an individual layer cannot fully hide a complete-layer transfer. MoE-CORE instead retains and prefetches selected experts, avoiding most of these transfers and yielding lower TPOT.

\subsection{MTP Compatibility and Best Performance}
Table~\ref{tab:mtp_performance} reports the MTP depth sweep for DeepSeek-V4-Flash-W4A8 on GSM8K at batch size 2 with AES enabled. Depth 2 achieves the lowest observed TPOT, 21.5 ms. As MTP depth increases from 1 to 3, the acceptance rate decreases from 95.21\% to 78.33\% and 60.78\%, while speculative H2D transfers increase. At depth 3, lower acceptance and additional transfer overhead offset the benefit of longer predictions, raising TPOT from 21.5 to 22.3 ms. Thus, depth 2 provides the best measured balance between prediction acceptance and transfer overhead on this platform.
% Source: Sheet2 G9:L9. The MTP-off layer-rate reference is F9.
% Report absolute TPOT: no matched MTP-off TPOT is recorded in this sweep.
\begin{table}[t]
\centering
\caption{MTP depth sweep on DeepSeek-V4-Flash-W4A8 with GSM8K, batch size 2, and AES enabled. Availability is the unweighted mean of 40 layer-wise effective expert-availability rates. Changes are in percentage points (pp) relative to the MTP-off mean of 99.972\%.}
\label{tab:mtp_performance}
\small
\setlength{\tabcolsep}{3pt}
\begin{tabular}{@{}crrrr@{}}
\toprule
\textbf{MTP} & \textbf{TPOT} & \textbf{Accept.} & \textbf{Avail.} & \(\boldsymbol{\Delta}\) \\
\textbf{depth} & \textbf{(ms)} & \textbf{(\%)} & \textbf{(\%)} & \textbf{(pp)} \\
\midrule
1 & 23.1 & 95.21 & 99.785 & \(-0.186\) \\
2 & \textbf{21.5} & 78.33 & 99.607 & \(-0.365\) \\
3 & 22.3 & 60.78 & 99.439 & \(-0.532\) \\
\bottomrule
\end{tabular}
\end{table}

Across MTP depths 1--3, effective expert availability exceeds 99.4\%; the largest decrease from the MTP-off reference is 0.532 percentage points. These measurements demonstrate compatibility with MTP and identify depth 2 as the best measured setting. Effective availability includes invocations served by resident substitutes rather than only exact hits of the originally selected experts.

\subsection{Ablation}
Unless otherwise stated, the ablations use GSM8K and DeepSeek-V4-Flash. Each experiment changes one component while staying within the same expert-memory budget and retaining the inference configuration.

\subsubsection{Effect of Expert-Cache Capacity}

\begin{figure}[t]
    \centering
    \includegraphics[width=\linewidth]{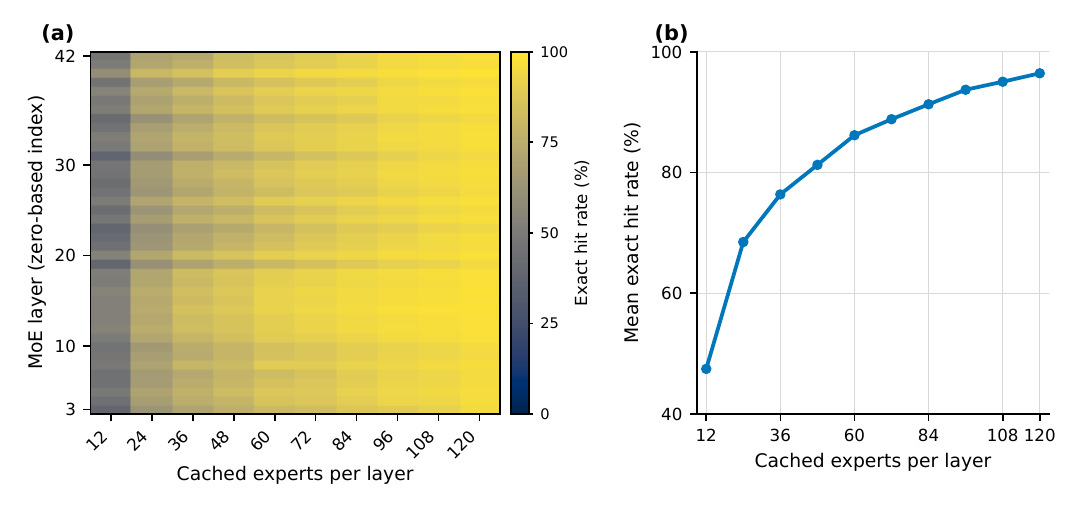}
    \caption{Effect of uniform expert-cache capacity on DeepSeek-V4-Flash-W4A8 with GSM8K. (a) Exact hit rate on each comparable layer (zero-based indices 3--42) for capacities of 12--120 experts per layer. (b) Unweighted mean exact hit rate over those 40 layers.}
    \label{fig:layer_cache_sensitivity}
\end{figure}

Figure~\ref{fig:layer_cache_sensitivity} evaluates the relationship between cache capacity and exact expert hit rate on the 40 comparable DeepSeek MoE layers (zero-based indices 3--42). Increasing the number of cached experts per layer from 12 to 120 raises the unweighted mean hit rate from 47.47\% to 96.43\%. Expanding from 12 to 24 experts gains 21.01 percentage points, whereas expanding from 84 to 120 gains only 5.13 points. Thus, the benefit of uniformly increasing cache capacity diminishes as more experts are resident.

\subsubsection{Effect of Decode-Cache Components}
\begin{figure}
    \centering
    \includegraphics[width=\linewidth]{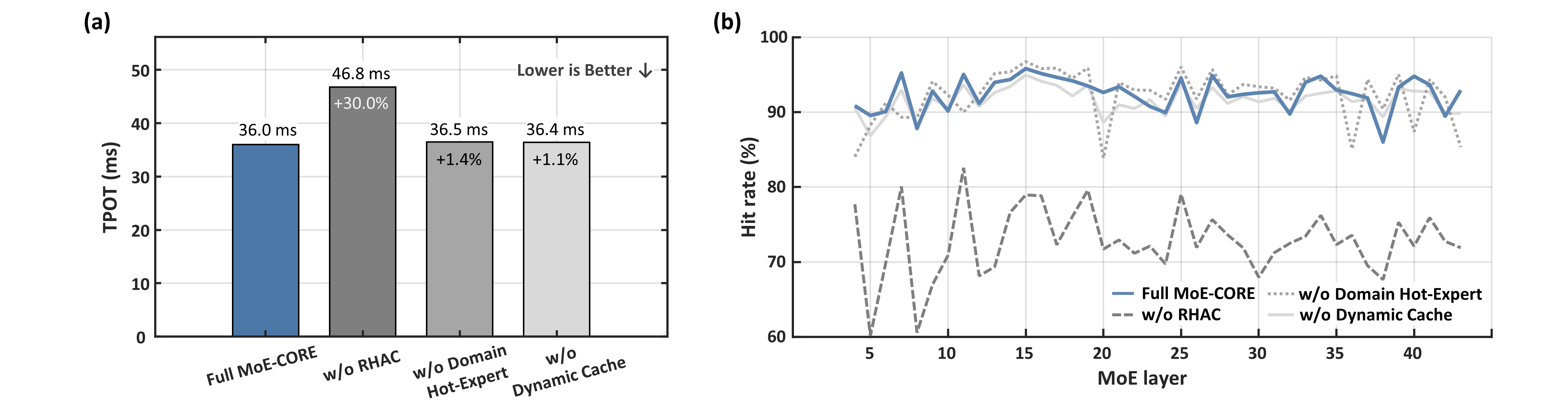}
    \caption{Ablation of the decode expert cache. (a) TPOT for the full system and each ablated configuration. (b) Per-layer exact cache hit rate under the same configurations. The ``w/o Dynamic Cache'' variant uses uniform per-layer capacity.}
    \label{fig:component_ablation}
\end{figure}

Figure~\ref{fig:component_ablation} compares the full configuration with three ablations: removing RHAC, removing domain-conditioned initialization, and replacing the nonuniform allocation with a uniform per-layer allocation. The full configuration's nonuniform capacity assignment was selected through repeated offline trials. Disabling RHAC increases TPOT from 36.0 to 46.8 ms, corresponding to a 30.0\% degradation. The exact expert hit rate simultaneously decreases from above 90\% in most layers to approximately 60--82\%. Because this reduction occurs throughout the model, RHAC is the principal mechanism for maintaining an effective resident working set during decode.

Removing the domain hot-expert profile increases TPOT to 36.5 ms (1.4\%), while replacing the nonuniform allocation with a uniform per-layer cache increases TPOT to 36.4 ms (1.1\%). Their layer-wise curves remain close to that of the full system, but exhibit localized reductions in hit rate. Domain profiles initialize cache contents, the selected nonuniform allocation gives low-hit layers more capacity, and RHAC adapts the resident set during decoding. The complete system combines these mechanisms and achieves the lowest TPOT.

\subsubsection{Effect of Cross-Layer Prefetch Depth}
\begin{figure}
    \centering
    \includegraphics[width=\linewidth]{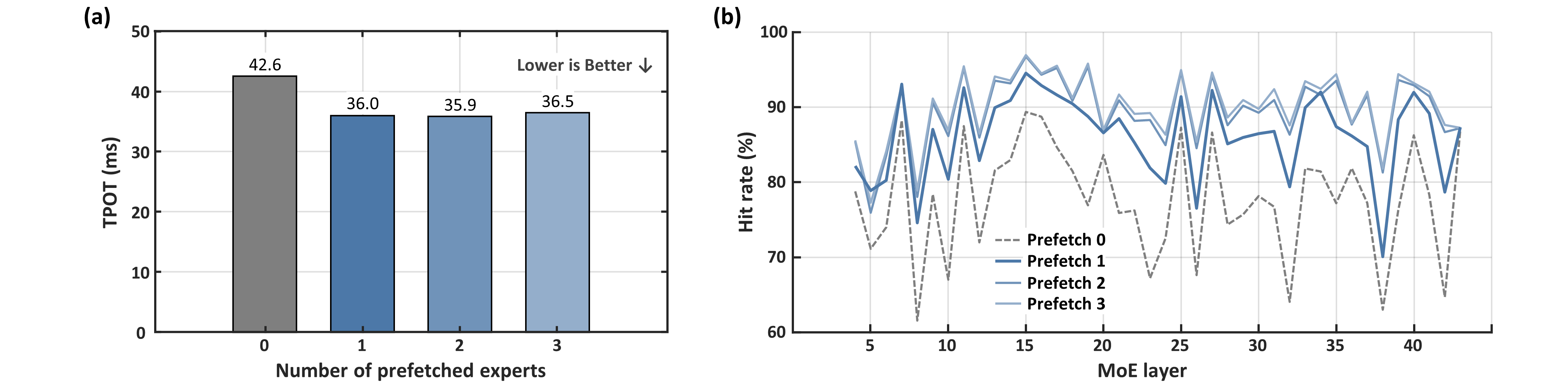}
    \caption{Effect of cross-layer prefetch depth. (a) TPOT for zero to three prefetched experts. (b) Per-layer exact cache hit rate at each depth.}
    \label{fig:prefetch_depth}
\end{figure}

Figure~\ref{fig:prefetch_depth} compares cross-layer prefetch depths from zero to three experts. Without prefetching, TPOT reaches 42.6 ms. Prefetching one expert reduces TPOT to 36.0 ms, an improvement of 15.5\%. Increasing the depth to two experts produces the minimum TPOT of 35.9 ms, only 0.1 ms below the one-expert configuration, whereas a depth of three increases TPOT to 36.5 ms.

Mean exact hit rate rises from 89.05\% without prefetching to 92.43\%, 93.81\%, and 94.05\% at depths one, two, and three, respectively. Beyond depth one, the additional hit-rate gains are small. At depth three, the H2D time for three predicted experts cannot be fully hidden by the limited per-layer computation during decode, so the additional transfers outweigh the marginal hit-rate gain and increase TPOT. The first predicted expert thus captures nearly all of the latency benefit. MoE-CORE uses a depth of one by default because it remains within 0.3\% of the minimum TPOT while issuing fewer speculative transfers.

\subsubsection{Effect of Approximate Expert Substitution}
Approximate expert substitution (AES) resolves an eligible residual miss with an unselected expert that is already resident on the NPU. We define \emph{effective expert availability} as the fraction of selected expert invocations that can proceed without a demand-triggered H2D transfer. With AES disabled, it reflects availability from caching and completed prefetches; with AES enabled, it also counts invocations served by resident substitutes.

\begin{figure}
    \centering
    \includegraphics[width=\linewidth]{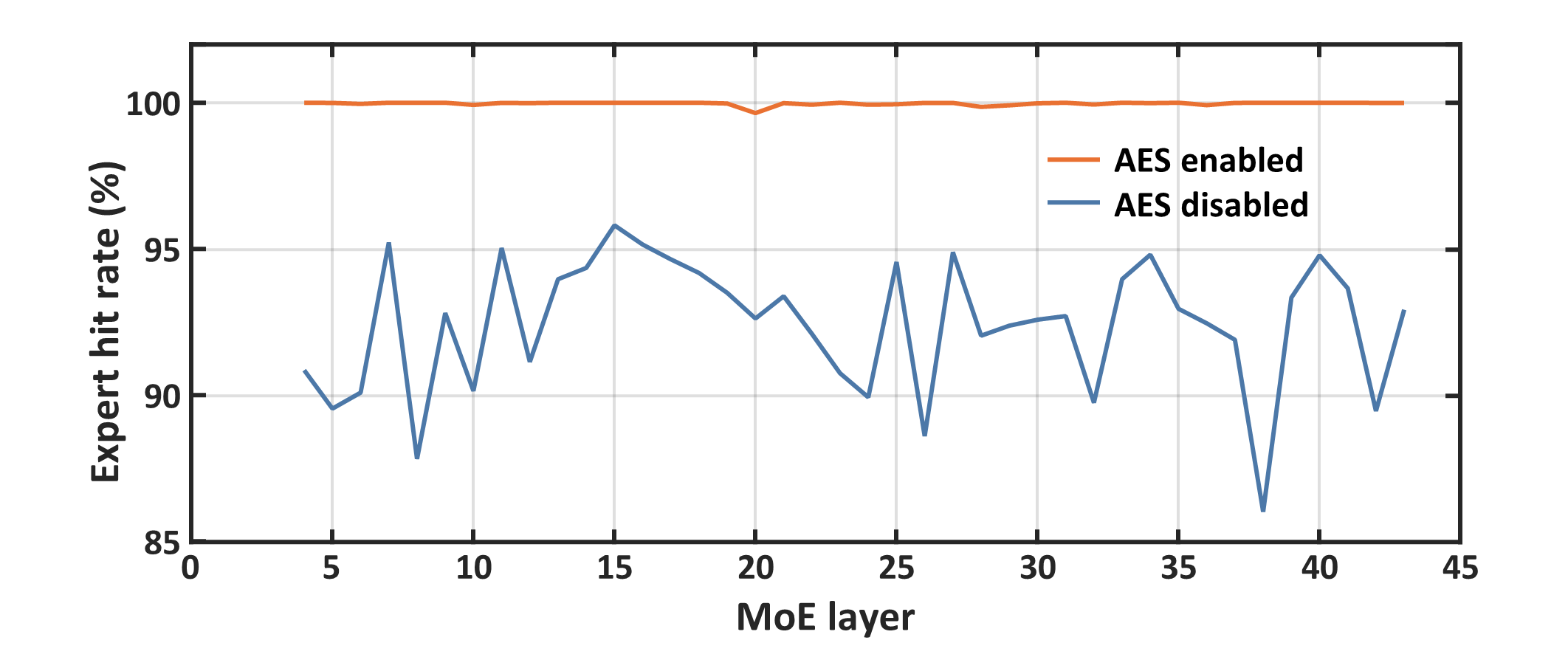}
    \caption{Expert availability with and without AES. The disabled curve measures availability of the original selected experts; the enabled curve additionally counts invocations served by resident substitutes.}
    \label{fig:aes_effect}
\end{figure}

The recorded mean over 40 layer rates rises from 92.43\% without AES to 99.972\% with AES. Figure~\ref{fig:aes_effect} visualizes this near-complete effective availability. This near-unity layer mean shows that, on average across the evaluated layers, MoE-CORE serves nearly all selected expert invocations without a demand-triggered H2D transfer.

\begin{table}[t]
\centering
\caption{Recorded AES latency under different decode configurations.}
\label{tab:aes_latency}
\small
\setlength{\tabcolsep}{4pt}
\begin{tabular}{@{}lccr@{}}
\toprule
\textbf{AES} & \textbf{Batch size} & \textbf{MTP depth}
& \textbf{TPOT (ms)} \\
\midrule
Disabled & 1 & Off & 36.0 \\
Enabled & 1 & Off & 37.7 \\
Enabled & 2 & 2 & 22.0 \\
\bottomrule
\end{tabular}
\end{table}

At batch size 1 with MTP disabled, enabling AES increases TPOT from 36.0 to 37.7 ms, a 4.7\% overhead, due to the CPU-side computation required for expert substitution.

\section{Related Work}
\paragraph{Expert offloading and heterogeneous execution.}
MoE-Infinity exploits activation traces to guide expert caching and prefetching, while FineMoE uses fine-grained routing patterns and prompt-derived hints for expert management \cite{xue2025moeinfinityefficientmoeinference,yu2025taminglatencymemorytradeoffmoebased}. Fiddler and KTransformers use host computation as well as device execution to accommodate experts under limited device memory \cite{kamahori2025fiddler,ktransformer}. HOBBIT uses lower-precision versions of less critical missed experts to reduce transfer cost \cite{tang2024hobbit}. MoE-CORE instead focuses on exact-expert residency and transfer scheduling within a single-NPU memory budget.

\paragraph{Transfer scheduling and routing-guided residency.}
MoE-Lightning pipelines host, device, and I/O work with paged weights, whereas Klotski uses an expert-aware multi-batch pipeline to increase computation--transfer overlap \cite{cao2024moelightning,fang2025klotski}. DuoServe-MoE explicitly separates prefill and decode scheduling and applies phase-specific expert prefetching \cite{zhang2026duoservemoedualphaseexpertprefetch}. SiDA-MoE and ProMoE train auxiliary predictors for proactive expert loading; Fate and AdapMoE exploit cross-layer routing information \cite{MLSYS2024_698cfaf7,song2025promoefastmoebasedllm,Fate2026Fang,Zhong_2024}. DoMoE additionally uses domain-specific routing tables to constrain semantic expert prediction \cite{mu2026domoe}. In contrast to predictor-centered designs, MoE-CORE uses a workload-domain profile to initialize cache contents, updates residency from observed routing history, and evaluates the original next-layer router early for prefetching.

\paragraph{Approximate execution and speculative decoding.}
SMoE and BuddyMoE examine expert substitution when selected experts are unavailable, while SpecMoEOff uses speculative decoding to enlarge expert workloads and hide offloading latency \cite{zhu2026smoealgorithmsystemcodesignpushing,wang2025buddymoeexploitingexpertredundancy,wang2025acceleratingmixtureofexpertsinferencehiding}. MoE-CORE's optional substitution is restricted to resident, unselected experts close to the selected expert's routing score at the Top-$k$ boundary; it does not assert functional equivalence from score proximity. Our MTP experiment assesses the compatibility of the expert-offloading path with multi-token prediction rather than introducing a new speculation algorithm. More broadly, long-context serving systems such as Sarathi-Serve and DistServe schedule prefill and decode to improve serving efficiency \cite{298679,zhong2024distservedisaggregatingprefilldecoding}; our focus is expert-weight residency and transfer within a single appliance.

\section{Conclusion}
MoE-CORE addresses expert-weight offloading for memory-constrained MoE inference on compact, single-NPU AI appliances. Measurements across five workloads reveal recurrent low-hit layers under uniform per-layer cache sizes and workload-dependent expert rankings; the examined routing-score distribution also illustrates a narrow gap near the Top-$k$ boundary. These observations motivate a runtime that separates prefill staging from decode-time expert management and coordinates nonuniform cache capacity, domain-informed initialization, routing-history-aware replacement, and cross-layer prefetching. Across five workloads per model and under the stated output caps, the main exact-execution configuration records mean TPOT of 38.0--44.8 ms on DeepSeek and 206.6--220.5 ms on GLM, versus 1268.9--1269.1 and 5941.5--5941.8 ms, respectively, for the evaluated vLLM Prefetch configurations. Under an 84-GB NPU-memory cap, a separate DeepSeek experiment with optional approximate substitution reaches the best measured TPOT of 21.5 ms at MTP depth 2. The results support coordinated residency and transfer scheduling as a practical means of reducing decode latency when the full MoE expert set cannot remain on the device.

\section*{Limitations}

MoE-CORE does not jointly optimize expert-weight offloading and KV-cache management, and its evaluation is limited to a single NPU; multi-NPU expert-weight placement remains future work. Further study of longer MTP depths may clarify when their additional H2D transfers can be offset by accepted predictions. We also plan to coordinate the currently separate prefill and decode expert caches across consecutive sessions. Experts retained by RHAC after one session may be reused in the next session's prefill, allowing layer-wise staging to transfer only experts that are not already resident. Finally, RHAC's CPU-side updates add latency; reducing this computation is another direction for optimization.

\bibliography{ref}

@misc{shazeer2017outrageouslylargeneuralnetworks,
      title={Outrageously Large Neural Networks: The Sparsely-Gated Mixture-of-Experts Layer}, 
      author={Noam Shazeer and Azalia Mirhoseini and Krzysztof Maziarz and Andy Davis and Quoc Le and Geoffrey Hinton and Jeff Dean},
      year={2017},
      eprint={1701.06538},
      archivePrefix={arXiv},
      primaryClass={cs.LG},
      url={https://arxiv.org/abs/1701.06538}, 
}

@misc{yi2025edgemoeempoweringsparselarge,
      title={EdgeMoE: Empowering Sparse Large Language Models on Mobile Devices}, 
      author={Rongjie Yi and Liwei Guo and Shiyun Wei and Ao Zhou and Shangguang Wang and Mengwei Xu},
      year={2025},
      eprint={2308.14352},
      archivePrefix={arXiv},
      primaryClass={cs.LG},
      url={https://arxiv.org/abs/2308.14352}, 
}

@misc{deepseekai2026deepseekv4,
      title={DeepSeek-V4: Towards Highly Efficient Million-Token Context Intelligence},
      author={DeepSeek-AI},
      year={2026},
      eprint={2606.19348},
      archivePrefix={arXiv},
      url={https://arxiv.org/abs/2606.19348},
}

@misc{glm5team2026glm5vibecodingagentic,
      title={GLM-5: from Vibe Coding to Agentic Engineering},
      author={{GLM-5 Team}},
      year={2026},
      eprint={2602.15763},
      archivePrefix={arXiv},
      primaryClass={cs.LG},
      url={https://arxiv.org/abs/2602.15763},
}

@misc{kimik3_2026,
  author  = {Kimi-Team},
  year = {2026},
  title   = {KIMI K3: Open Frontier Intelligence},
  url     = {https://github.com/MoonshotAI/Kimi-K3/blob/main/k3_tech_report.pdf},
  urldate = {2026-07-28},
}

@misc{xue2025moeinfinityefficientmoeinference,
      title={MoE-Infinity: Efficient MoE Inference on Personal Machines with Sparsity-Aware Expert Cache}, 
      author={Leyang Xue and Yao Fu and Zhan Lu and Luo Mai and Mahesh Marina},
      year={2025},
      eprint={2401.14361},
      archivePrefix={arXiv},
      primaryClass={cs.LG},
      url={https://arxiv.org/abs/2401.14361}, 
}

@misc{yu2025taminglatencymemorytradeoffmoebased,
      title={Taming Latency-Memory Trade-Off in MoE-Based LLM Serving via Fine-Grained Expert Offloading}, 
      author={Hanfei Yu and Xingqi Cui and Hong Zhang and Hao Wang and Hao Wang},
      year={2025},
      eprint={2502.05370},
      archivePrefix={arXiv},
      primaryClass={cs.LG},
      url={https://arxiv.org/abs/2502.05370}, 
}

@inproceedings{MLSYS2024_698cfaf7,
 author = {Du, Zhixu and Li, Shiyu and Wu, Yuhao and Jiang, Xiangyu and Sun, Jingwei and Zheng, Qilin and Wu, Yongkai and Li, Ang and Li, Hai Helen and Chen, Yiran},
 booktitle = {Proceedings of Machine Learning and Systems},
 editor = {P. Gibbons and G. Pekhimenko and C. De Sa},
 pages = {224--238},
 title = {SiDA: Sparsity-Inspired Data-Aware Serving for Efficient and Scalable  Large Mixture-of-Experts Models},
 url = {https://proceedings.mlsys.org/paper_files/paper/2024/file/698cfaf72a208aef2e78bcac55b74328-Paper-Conference.pdf},
 volume = {6},
 year = {2024}
}

@misc{song2025promoefastmoebasedllm,
      title={ProMoE: Fast MoE-based LLM Serving using Proactive Caching}, 
      author={Xiaoniu Song and Zihang Zhong and Rong Chen and Haibo Chen},
      year={2025},
      eprint={2410.22134},
      archivePrefix={arXiv},
      primaryClass={cs.DC},
      url={https://arxiv.org/abs/2410.22134}, 
}

@inproceedings{Fate2026Fang,
author = {Fang, Zhiyuan and Yu, Xingfan and Huang, Yuegui and Hong, Zicong and Lyu, Yufeng and Chen, Wuhui and Yu, Yue and Yu, Fan},
title = {Fate: Fast Edge Inference of Mixture-of-Experts Models via Cross-Layer Gate},
year = {2026},
isbn = {9798400723070},
publisher = {Association for Computing Machinery},
address = {New York, NY, USA},
url = {https://doi.org/10.1145/3774904.3792527},
doi = {10.1145/3774904.3792527},
booktitle = {Proceedings of the ACM Web Conference 2026},
pages = {5537–5546},
numpages = {10},
location = {United Arab Emirates},
series = {WWW '26}
}

@inproceedings{Zhong_2024, series={ICCAD ’24},
   title={AdapMoE: Adaptive Sensitivity-based Expert Gating and Management for Efficient MoE Inference},
   url={http://dx.doi.org/10.1145/3676536.3676741},
   DOI={10.1145/3676536.3676741},
   booktitle={Proceedings of the 43rd IEEE/ACM International Conference on Computer-Aided Design},
   publisher={ACM},
   author={Zhong, Shuzhang and Liang, Ling and Wang, Yuan and Wang, Runsheng and Huang, Ru and Li, Meng},
   year={2024},
   month=Oct, pages={1–9},
   collection={ICCAD ’24} }

@misc{zhang2026duoservemoedualphaseexpertprefetch,
      title={DuoServe-MoE: Dual-Phase Expert Prefetch and Caching for LLM Inference QoS Assurance}, 
      author={Yuning Zhang and Grant Pinkert and Nan Yang and Yanli Li and Dong Yuan},
      year={2026},
      eprint={2509.07379},
      archivePrefix={arXiv},
      primaryClass={cs.DC},
      url={https://arxiv.org/abs/2509.07379}, 
}

@misc{zhu2026smoealgorithmsystemcodesignpushing,
      title={SMoE: An Algorithm-System Co-Design for Pushing MoE to the Edge via Expert Substitution}, 
      author={Guoying Zhu and Meng Li and Haipeng Dai and Xuechen Liu and Weijun Wang and Keran Li and Jun xiao and Ligeng Chen and Wei Wang},
      year={2026},
      eprint={2508.18983},
      archivePrefix={arXiv},
      primaryClass={cs.AI},
      url={https://arxiv.org/abs/2508.18983}, 
}

@misc{wang2025buddymoeexploitingexpertredundancy,
      title={BuddyMoE: Exploiting Expert Redundancy to Accelerate Memory-Constrained Mixture-of-Experts Inference}, 
      author={Yun Wang and Lingyun Yang and Senhao Yu and Yixiao Wang and Ruixing Li and Zhixiang Wei and James Yen and Zhengwei Qi},
      year={2025},
      eprint={2511.10054},
      archivePrefix={arXiv},
      primaryClass={cs.LG},
      url={https://arxiv.org/abs/2511.10054}, 
}

@misc{fang2026agentlongbenchcontrollablelongbenchmark,
      title={AgentLongBench: A Controllable Long Benchmark For Long-Contexts Agents via Environment Rollouts}, 
      author={Shicheng Fang and Yuxin Wang and Xiaoran Liu and Jiahao Lu and Chuanyuan Tan and Xinchi Chen and Yining Zheng and Xuanjing Huang and Xipeng Qiu},
      year={2026},
      eprint={2601.20730},
      archivePrefix={arXiv},
      primaryClass={cs.CL},
      url={https://arxiv.org/abs/2601.20730}, 
}

@inproceedings {298679,
author = {Amey Agrawal and Nitin Kedia and Ashish Panwar and Jayashree Mohan and Nipun Kwatra and Bhargav Gulavani and Alexey Tumanov and Ramachandran Ramjee},
title = {Taming {Throughput-Latency} Tradeoff in {LLM} Inference with {Sarathi-Serve}},
booktitle = {18th USENIX Symposium on Operating Systems Design and Implementation (OSDI 24)},
year = {2024},
isbn = {978-1-939133-40-3},
address = {Santa Clara, CA},
pages = {117--134},
url = {https://www.usenix.org/conference/osdi24/presentation/agrawal},
publisher = {USENIX Association},
month = jul
}

@misc{zhong2024distservedisaggregatingprefilldecoding,
      title={DistServe: Disaggregating Prefill and Decoding for Goodput-optimized Large Language Model Serving}, 
      author={Yinmin Zhong and Shengyu Liu and Junda Chen and Jianbo Hu and Yibo Zhu and Xuanzhe Liu and Xin Jin and Hao Zhang},
      year={2024},
      eprint={2401.09670},
      archivePrefix={arXiv},
      primaryClass={cs.DC},
      url={https://arxiv.org/abs/2401.09670}, 
}

@misc{vllm_prefetch_offloader,
  author = {{vLLM Project}},
  title = {Prefetch-based CPU Offloading in vLLM},
  year = {2026},
  howpublished = {Source code in the vLLM repository},
  url = {https://github.com/vllm-project/vllm/blob/main/vllm/model_executor/offloader/prefetch.py},
  note = {Accessed 2026-09-30},
}

@inproceedings{ICLR2024_d3cf1559,
 author = {Gou, Zhibin and Shao, Zhihong and Gong, Yeyun and shen, yelong and Yang, Yujiu and Huang, Minlie and Duan, Nan and Chen, Weizhu},
 booktitle = {International Conference on Learning Representations},
 editor = {B. Kim and Y. Yue and S. Chaudhuri and K. Fragkiadaki and M. Khan and Y. Sun},
 pages = {48362--48395},
 title = {ToRA: A Tool-Integrated Reasoning Agent for Mathematical Problem Solving},
 url = {https://proceedings.iclr.cc/paper_files/paper/2024/file/d3cf1559a8795eb1ed2b3ad52409ac7d-Paper-Conference.pdf},
 volume = {2024},
 year = {2024}
}

@inproceedings{jimenez2024swe,
  title={Swe-bench: Can language models resolve real-world github issues?},
  author={Jimenez, Carlos E and Yang, John and Wettig, Alexander and Yao, Shunyu and Pei, Kexin and Press, Ofir and Narasimhan, Karthik},
  booktitle={International Conference on Learning Representations},
  volume={2024},
  pages={54107--54157},
  year={2024}
}

@article{yao2024tau,
  title={tau-bench: A Benchmark for Tool-Agent-User Interaction in Real-World Domains},
  author={Yao, Shunyu and Shinn, Noah and Razavi, Pedram and Narasimhan, Karthik},
  journal={arXiv preprint arXiv:2406.12045},
  year={2024}
}

@misc{wang2025acceleratingmixtureofexpertsinferencehiding,
      title={Accelerating Mixture-of-Experts Inference by Hiding Offloading Latency with Speculative Decoding}, 
      author={Zhibin Wang and Zhonghui Zhang and Yuhang Zhou and Zibo Wang and Mo Zhou and Peng Jiang and Weilin Cai and Chengying Huan and Rong Gu and Sheng Zhong and Chen Tian},
      year={2025},
      eprint={2508.21706},
      archivePrefix={arXiv},
      primaryClass={cs.DC},
      url={https://arxiv.org/abs/2508.21706}, 
}

@misc{brown1956exponential,
  author       = {Brown, Robert G.},
  title        = {Exponential Smoothing for Predicting Demand},
  year         = {1956},
  howpublished = {Presented at the Tenth National Meeting of the Operations Research Society of America},
  address      = {San Francisco, CA, USA},
  note         = {Also issued as a monograph by Arthur D. Little, Inc.}
}

@misc{cobbe2021trainingverifierssolvemath,
      title={Training Verifiers to Solve Math Word Problems}, 
      author={Karl Cobbe and Vineet Kosaraju and Mohammad Bavarian and Mark Chen and Heewoo Jun and Lukasz Kaiser and Matthias Plappert and Jerry Tworek and Jacob Hilton and Reiichiro Nakano and Christopher Hesse and John Schulman},
      year={2021},
      eprint={2110.14168},
      archivePrefix={arXiv},
      primaryClass={cs.LG},
      url={https://arxiv.org/abs/2110.14168}, 
}

@misc{dekoninck2026matharena,
      title={Beyond Benchmarks: MathArena as an Evaluation Platform for Mathematics with LLMs}, 
      author={Jasper Dekoninck and Nikola Jovanović and Tim Gehrunger and Kári Rögnvaldsson and Ivo Petrov and Chenhao Sun and Martin Vechev},
      year={2026},
      eprint={2605.00674},
      archivePrefix={arXiv},
      primaryClass={cs.CL},
      url={https://arxiv.org/abs/2605.00674}, 
}

@misc{chen2021evaluatinglargelanguagemodels,
      title={Evaluating Large Language Models Trained on Code}, 
      author={Mark Chen and Jerry Tworek and Heewoo Jun and Qiming Yuan and Henrique Ponde de Oliveira Pinto and Jared Kaplan and Harri Edwards and Yuri Burda and Nicholas Joseph and Greg Brockman and Alex Ray and Raul Puri and Gretchen Krueger and Michael Petrov and Heidy Khlaaf and Girish Sastry and Pamela Mishkin and Brooke Chan and Scott Gray and Nick Ryder and Mikhail Pavlov and Alethea Power and Lukasz Kaiser and Mohammad Bavarian and Clemens Winter and Philippe Tillet and Felipe Petroski Such and Dave Cummings and Matthias Plappert and Fotios Chantzis and Elizabeth Barnes and Ariel Herbert-Voss and William Hebgen Guss and Alex Nichol and Alex Paino and Nikolas Tezak and Jie Tang and Igor Babuschkin and Suchir Balaji and Shantanu Jain and William Saunders and Christopher Hesse and Andrew N. Carr and Jan Leike and Josh Achiam and Vedant Misra and Evan Morikawa and Alec Radford and Matthew Knight and Miles Brundage and Mira Murati and Katie Mayer and Peter Welinder and Bob McGrew and Dario Amodei and Sam McCandlish and Ilya Sutskever and Wojciech Zaremba},
      year={2021},
      eprint={2107.03374},
      archivePrefix={arXiv},
      primaryClass={cs.LG},
      url={https://arxiv.org/abs/2107.03374}, 
}

@misc{hendrycks2021measuringmassivemultitasklanguage,
      title={Measuring Massive Multitask Language Understanding}, 
      author={Dan Hendrycks and Collin Burns and Steven Basart and Andy Zou and Mantas Mazeika and Dawn Song and Jacob Steinhardt},
      year={2021},
      eprint={2009.03300},
      archivePrefix={arXiv},
      primaryClass={cs.CY},
      url={https://arxiv.org/abs/2009.03300}, 
}

@misc{bai2025longbenchv2deeperunderstanding,
      title={LongBench v2: Towards Deeper Understanding and Reasoning on Realistic Long-context Multitasks}, 
      author={Yushi Bai and Shangqing Tu and Jiajie Zhang and Hao Peng and Xiaozhi Wang and Xin Lv and Shulin Cao and Jiazheng Xu and Lei Hou and Yuxiao Dong and Jie Tang and Juanzi Li},
      year={2025},
      eprint={2412.15204},
      archivePrefix={arXiv},
      primaryClass={cs.CL},
      url={https://arxiv.org/abs/2412.15204}, 
}

@inproceedings{ktransformer,
author = {Chen, Hongtao and Xie, Weiyu and Zhang, Boxin and Tang, Jingqi and Wang, Jiahao and Dong, Jianwei and Chen, Shaoyuan and Yuan, Ziwei and Lin, Chen and Qiu, Chengyu and Zhu, Yuening and Ou, Qingliang and Liao, Jiaqi and Chen, Xianglin and Ai, Zhiyuan and Wu, Yongwei and Zhang, Mingxing},
title = {KTransformers: Unleashing the Full Potential of CPU/GPU Hybrid Inference for MoE Models},
year = {2025},
isbn = {9798400718700},
publisher = {Association for Computing Machinery},
address = {New York, NY, USA},
url = {https://doi.org/10.1145/3731569.3764843},
doi = {10.1145/3731569.3764843},
booktitle = {Proceedings of the ACM SIGOPS 31st Symposium on Operating Systems Principles},
pages = {1014–1029},
numpages = {16},
location = {Lotte Hotel World, Seoul, Republic of Korea},
series = {SOSP '25}
}

@inproceedings{kamahori2025fiddler,
  title = {Fiddler: CPU-GPU Orchestration for Fast Inference of Mixture-of-Experts Models},
  author = {Keisuke Kamahori and Tian Tang and Yile Gu and Kan Zhu and Baris Kasikci},
  booktitle = {International Conference on Learning Representations},
  year = {2025},
  url = {https://arxiv.org/abs/2402.07033}
}

@misc{tang2024hobbit,
  title = {HOBBIT: A Mixed Precision Expert Offloading System for Fast MoE Inference},
  author = {Peng Tang and Jiacheng Liu and Xiaofeng Hou and Yifei Pu and Jing Wang and Pheng-Ann Heng and Chao Li and Minyi Guo},
  year = {2024},
  eprint = {2411.01433},
  archivePrefix = {arXiv},
  url = {https://arxiv.org/abs/2411.01433}
}

@misc{cao2024moelightning,
  title = {MoE-Lightning: High-Throughput MoE Inference on Memory-constrained GPUs},
  author = {Shiyi Cao and Shu Liu and Tyler Griggs and Peter Schafhalter and Xiaoxuan Liu and Ying Sheng and Joseph E. Gonzalez and Matei Zaharia and Ion Stoica},
  year = {2024},
  eprint = {2411.11217},
  archivePrefix = {arXiv},
  url = {https://arxiv.org/abs/2411.11217}
}

@misc{fang2025klotski,
  title = {Klotski: Efficient Mixture-of-Expert Inference via Expert-Aware Multi-Batch Pipeline},
  author = {Zhiyuan Fang and Yuegui Huang and Zicong Hong and Yufeng Lyu and Wuhui Chen and Yue Yu and Fan Yu and Zibin Zheng},
  year = {2025},
  eprint = {2502.06888},
  archivePrefix = {arXiv},
  url = {https://arxiv.org/abs/2502.06888}
}

@inproceedings{mu2026domoe,
  title = {DoMoE: Domain-Aware Semantic Expert Prediction for Efficient MoE Inference Under Expert Offloading},
  author = {Yao Mu and Fahao Chen and Wenbin Zhu and Mengying Zhao and Zhaoyan Shen and Dongxiao Yu},
  booktitle = {Proceedings of the Thirty-Fifth International Joint Conference on Artificial Intelligence},
  pages = {5901--5909},
  year = {2026},
  doi = {10.24963/ijcai.2026/657},
  url = {https://www.ijcai.org/proceedings/2026/657}
}

\end{document}